\documentclass{amsart}
\usepackage[a4paper,margin=1in]{geometry}
\usepackage{amssymb}
\usepackage[hyphens]{url}
\usepackage{amsmath}
\usepackage{amsthm}
\usepackage[title]{appendix}
\usepackage[utf8]{inputenc}
\usepackage[english]{babel}
\usepackage[authoryear,round]{natbib}
\usepackage{aliascnt}
\usepackage{hyperref}
\usepackage[capitalise,noabbrev]{cleveref}
\usepackage[section]{placeins}
\usepackage{enumitem}
\usepackage{caption}
\usepackage{float}
\usepackage{mathtools}
\usepackage{xcolor}
\usepackage{makecell}
\usepackage{booktabs}
\usepackage{nccmath}
\usepackage{tikz}
\usepackage{pgfplots}
\pgfplotsset{compat=1.18}

\hypersetup{
    colorlinks=true,
    linkcolor=blue,
    citecolor=teal,
    urlcolor=red
}

\hypersetup{linktoc=all}
\numberwithin{equation}{section}

\newtheorem{thm}{Theorem}[section]

\newaliascnt{lem}{thm}
\newtheorem{lem}[lem]{Lemma}
\aliascntresetthe{lem}
\newaliascnt{prop}{thm}
\newtheorem{prop}[prop]{Proposition}
\aliascntresetthe{prop}
\newaliascnt{cor}{thm}

\aliascntresetthe{cor}
\newaliascnt{obs}{thm}

\aliascntresetthe{obs}
\newaliascnt{defn}{thm}
\newtheorem{defn}[defn]{Definition}
\aliascntresetthe{defn}
\newaliascnt{assump}{thm}

\aliascntresetthe{assump}

\newaliascnt{con}{thm}

\aliascntresetthe{con}
\newaliascnt{que}{thm}

\aliascntresetthe{que}
\newaliascnt{prob}{thm}
\newtheorem{prob}[prob]{Problem}
\aliascntresetthe{prob}
\newaliascnt{fct}{thm}

\aliascntresetthe{fct}
\newtheorem{mainthm}{Theorem}

\crefname{mainthm}{theorem}{theorems}
\Crefname{mainthm}{Theorem}{Theorems}
\crefname{maincor}{corollary}{corollaries}
\Crefname{maincor}{Corollary}{Corollaries}
\newaliascnt{claim}{thm}

\aliascntresetthe{claim}
\theoremstyle{definition}
\newaliascnt{exm}{thm}
\newtheorem{exm}[exm]{Example}
\aliascntresetthe{exm}
\newaliascnt{exs}{thm}

\aliascntresetthe{exs}
\theoremstyle{remark}
\newaliascnt{rem}{thm}

\aliascntresetthe{rem}

\crefname{thm}{theorem}{theorems}
\Crefname{thm}{Theorem}{Theorems}
\crefname{lem}{lemma}{lemmas}
\Crefname{lem}{Lemma}{Lemmas}
\crefname{prop}{proposition}{propositions}
\Crefname{prop}{Proposition}{Propositions}
\crefname{cor}{corollary}{corollaries}
\Crefname{cor}{Corollary}{Corollaries}
\crefname{obs}{observation}{observations}
\Crefname{obs}{Observation}{Observations}
\crefname{defn}{definition}{definitions}
\Crefname{defn}{Definition}{Definitions}
\crefname{que}{question}{questions}
\Crefname{que}{Question}{Questions}
\crefname{prob}{problem}{problems}
\Crefname{prob}{Problem}{Problems}
\crefname{fct}{fact}{facts}
\Crefname{fct}{Fact}{Facts}
\crefname{claim}{claim}{claims}
\Crefname{claim}{Claim}{Claims}
\crefname{rem}{remark}{remarks}
\Crefname{rem}{Remark}{Remarks}
\crefname{exm}{example}{examples}
\Crefname{exm}{Example}{Examples}
\crefname{exs}{examples}{examples}
\Crefname{exs}{Examples}{Examples}
\crefname{figure}{figure}{figure}
\Crefname{figure}{Figure}{Figure}

\makeatletter
\def\@settitle{%
 \begin{center}%
 \baselineskip18\p@\relax
 \normalfont\Large\scshape
 \@title
 \end{center}%
}
\makeatother

\title[Allocation under unknown overlaps]{Allocation under unknown overlaps:\\
from resource consistency to maximum entropy}

\author{Michael Abraham}
\address{Institute for Advanced Torah Studies, Bar-Ilan University, Israel}
\email{mikyab@gmail.com}

\author{Nachi Avraham-Re'em}
\address{Faculty of Mathematics, Technion -- Israel Institute of Technology}
\email{nachi.avraham@gmail.com}


\begin{document}

\begin{abstract}
A divisible resource is associated with several agents. Often only the individual proportions are known but not their overlaps, as in O'Neill's (1982) simple claims problem, credit attribution, and the museum-pass problem. How should the resource be divided based on the individual proportions, when the full-information benchmark divides each unit equally among its associated agents?

\smallskip

In bankruptcy problems, Aumann and Maschler (1985) derived the Talmud rule by bilateral consistency, which reduces the population. In allocation under unknown overlaps problems, we propose \emph{resource consistency}, which reduces the resource: each association cuts the resource into two parts, and the other associations are reconstructed proportionally on each part while preserving their total sizes. The allocations on the two parts, weighted by their sizes, must reproduce the original allocation.

\smallskip

We prove that there exists a unique efficient, symmetric, null rule that satisfies resource consistency. It is the Shapley value of the game induced by the admissible overlap model of maximum entropy, with an explicit formula for the allocation in terms of the \emph{overlap factor}, calculated as a root of a polynomial. When some agent has total association, the rule coincides with Diskin’s (1902) division and O’Neill’s (1982) method of random claims. The maximum-entropy model preserves the prescribed proportions, and thus it repairs the flaw identified by O'Neill in the method of random claims.
\end{abstract}

\maketitle

\tableofcontents

\section{Introduction and results}

Consider a divisible resource with which several agents are associated, and suppose that what is known is the size of the individual associations of the agents, as proportions of the resource, but not which parts these are. The parts associated with different agents may then overlap, and the overlaps, namely the proportions of the resource associated with several agents, are unknown. How should the resource be divided on the basis of the individual proportions alone?

\smallskip

A fundamental instance of an allocation under unknown overlaps problem is contested ownership, originating in an ancient Talmudic problem \citep[see][]{Thomson2019}. A person claims to own a garment and two others each claim to own half of it, so the individual claim proportions are \(\left(1,1/2,1/2\right)\). When each part is divided equally among its claimants, the awards depend on how the two half-claims overlap. If they concern the same half, the division is \(\left(2/3,1/6,1/6\right)\), while if they concern disjoint halves, the division is \(\left(1/2,1/4,1/4\right)\).

\smallskip

Other common instances of the allocation under unknown overlaps problem include the following.
\begin{itemize}
    \item The revenue from museum passes at a common price is divided among the participating museums. Each pass is used at least once and entitles its holder to one visit per museum.\footnote{The \href{https://www.parismuseumpass.fr/en/conditions-vente}{Paris Museum Pass}, for example, allows one visit to each participating site.} The only information about visits available for the division is the number of visits at each museum.
    \item Sales are credited to advertising channels. Each sale under consideration carries one unit of credit, and its buyer was exposed to at least one channel. For each channel, only the number of such sales with a recorded exposure through that channel before the sale is known.
    \item The editing credit for a film is divided among its editors, with equal weight given to each second of footage. Several editors may work on the same part, and only the total duration of the footage on which each editor worked is known.
    \item The revenue from a subscription bundle at a common price is divided among its content providers. Each subscriber selects one or more providers to include in the subscription, and for each provider only the number of subscribers who selected it is available for the division.
\end{itemize}

\smallskip

Related works on some of these problems are discussed in \S\ref{sct:literature}. We take as our full-information benchmark equal division of each unit among its associate agents, which is the Shapley value of an appropriate coalitional game (\S\ref{sct:alloc}). Let us formulate the problem generally.

\begin{prob}\label{prob:overlaps}
A divisible resource of size \(1\) is associated with \(N\) agents, each unit of the resource with at least one of them. The \(i^{\mathrm{th}}\) agent is associated with a proportion \(w_{i}\) of the resource and \(\sum\nolimits_{i=1}^{N}w_{i}\geq1\). The overlaps are unknown. How should the resource be shared, where equal division of each unit among its associated agents is the full-information benchmark?
\end{prob}

The \emph{simple claims problem} was formulated by \citet[\S2]{ONeill1982} as an extension of the Talmudic \emph{contested garment problem} to \(N\) claimants. We view it as an instance of \cref{prob:overlaps}.

\begin{prob}[Contested ownership]\label{prob:Ncontgarm}
\(N\) people make ownership claims to a divisible object of size \(1\). The \(i^{\mathrm{th}}\) person claims a proportion \(w_{i}\) of the object and \(\sum\nolimits_{i=1}^{N}w_{i}\geq1\). How should the object be divided? We assume that every part of the object is claimed by some claimant.
\end{prob}

For \(N=2\) agents, the individual proportions determine the overlap size as \(w_{1}+w_{2}-1\), so each agent receives the part associated with him alone, and the overlap is divided equally, which gives
\[\operatorname{CD}\left(w_{1},w_{2}\right)\coloneq\Big(1-w_{2}+\mfrac{1-\left(1-w_{1}+1-w_{2}\right)}{2},1-w_{1}+\mfrac{1-\left(1-w_{1}+1-w_{2}\right)}{2}\Big)=\Big(\mfrac{1+w_{1}-w_{2}}{2},\mfrac{1+w_{2}-w_{1}}{2}\Big).\]
This is the \emph{concede-and-divide} rule, also called the \emph{contested garment method} \citep[\S2.3]{Moulin2003}.

\smallskip

For \(N\geq3\) agents the proportions generally do not determine the overlaps, as the opening garment example shows, and therefore extending concede-and-divide requires an additional principle. For bankruptcy problems, \citet{AumannMaschler1985} showed that the \emph{Talmud rule} is the unique extension of concede-and-divide that is \emph{bilaterally consistent}, that is, whenever all agents but two leave with their shares, it assigns the same shares to the remaining two in the reduced problem. For \cref{prob:overlaps}, the Talmud rule fails strict population and strict cross-association monotonicity (\cref{exm:diskin}). It also fails compatibility with the full-information benchmark; indeed, for claims vector \(\left(1,1/2,1\right)\), every unit of the second claimant's half is claimed by all three, so equal division gives him \(1/6\) under any admissible overlap model. The Talmud rule, however, awards the second claimant \(1/4\) (Appendix~\ref{app:suppwt}).

\smallskip

There is a fundamental distinction between bankruptcy problems and contested ownership, which was already mentioned by \citet[\S2, p.~199]{AumannMaschler1985}, \citet[p.~51, n.~1]{Dagan1996}, \citet[\S1.3, \S15.1]{Thomson2019}.\footnote{Aumann and Maschler's discussion about the Talmud rule includes the contested garment problem, but their motivation was an enigmatic Talmudic discussion about the \emph{three wives problem}, which is a bankruptcy problem per se.} In bankruptcy problems, the claims are debts that can all be valid simultaneously, and the conflict is over insufficient material to pay them all. Thus, the different debts do not overlap each other in a way affecting the rights. In contested ownership, by contrast, the conflict is over incompatible assertions of ownership which may concern the same part of the object. The awards under the full-information benchmark can then depend on how their claims overlap.

\subsection{Resource consistency}\label{sct:consistency-intro}

Consistency axioms compare the awards in a problem with those obtained by applying the same rule to reduced problems; see  \citet[\S2]{Thomson2012}. In bilateral consistency, the reduced problem is given by each pair of agents. For \cref{prob:overlaps}, we consider reductions of the resource instead. An agent's association cuts the resource into two parts, one has the size of his association and the other has the complement size. Under the full-information benchmark, the weighted mixture of the two separate allocations reproduces the original awards \eqref{eq:benchmark-cut}.

\smallskip

With unknown overlaps, the sizes of the two parts are known, but the other associations within each of them must be reconstructed. On each part, we preserve the relative sizes of these associations, while preserving each agent's total association. The original awards need to be obtainable as a weighted mixture of the awards on the two parts.

\smallskip

Denote \(\left[N\right]\coloneq\left\{1,\dotsc,N\right\}\). For \(N\geq1\), we work on the domain
\[\Omega_{N}\coloneq\big\{\vec{w}\in\left[0,1\right]^{N}:\sum\nolimits_{i=1}^{N}w_{i}\geq1\big\}.\]
A \emph{rule} is a family of maps \(\left(R_{N}\right)_{N\geq1}\) of the form \(R_{N}:\Omega_{N}\longrightarrow\left[0,1\right]^{N}\). We take as a convention that \(R_{1}\left(1\right)=1\). For \(\vec{w}\in\Omega_{N}\) and \(j\in\left[N\right]\), write \(\vec{w}_{-j}\) for the vector obtained by deleting the \(j^{\mathrm{th}}\) coordinate, and \(\left(a,\vec{w}_{-j}\right)\) for the vector obtained by inserting \(a\) in the \(j^{\mathrm{th}}\) coordinate.

\smallskip

For an agent \(j\) with \(0<w_{j}<1\), the two parts have sizes \(w_{j}\) and \(1-w_{j}\). A proportional reconstruction gives them the normalized association vectors
\[\left(1,c_{j}\vec{w}_{-j}\right)
\quad\text{and}\quad
\left(0,c_{j}^{\ast}\vec{w}_{-j}\right).\]
Preserving each original association requires
\(w_{j}c_{j}w_{i}+\left(1-w_{j}\right)c_{j}^{\ast}w_{i}=w_{i}\), so \(w_{j}c_{j}+\left(1-w_{j}\right)c_{j}^{\ast}=1\) and
\begin{equation}\label{eq:cjast}
c_{j}^{\ast}=\mfrac{1-c_{j}w_{j}}{1-w_{j}}.
\end{equation}
A factor \(c_{j}\geq0\) is \emph{admissible} if both reconstructed vectors \(\left(1,c_{j}\vec{w}_{-j}\right)\) and \(\left(0,c_{j}^{\ast}\vec{w}_{-j}\right)\) belong to \(\Omega_{N}\).

\begin{defn}[Resource consistency]\label{dfn:resource-consistency}
A rule \(\left(R_{N}\right)_{N\geq1}\) is \emph{resource consistent} if for every \(\vec{w}\in\Omega_{N}\) and every \(j\in\left[N\right]\) with \(0<w_{j}<1\), there is an admissible factor \(c_{j}\) such that
\begin{equation}\label{eq:resource-consistency}
R_{N}\left(\vec{w}\right)=w_{j}R_{N}\left(1,c_{j}\vec{w}_{-j}\right)+\left(1-w_{j}\right)R_{N}\left(0,c_{j}^{\ast}\vec{w}_{-j}\right).
\end{equation}
\end{defn}

While the definition of resource consistency does not specify the admissible factors \(c_{j}\), we prove that they are uniquely determined.

\smallskip

On the total association domain, every admissible factor necessarily equals \(1\). Indeed, fix \(k\) with \(w_{k}=1\). For every \(j\) with \(0<w_{j}<1\), admissibility requires \(c_{j}\leq1\) and \(c_{j}^{\ast}\leq1\). By the definition \eqref{eq:cjast}, the latter inequality gives \(c_{j}\geq1\). Thus, on the total association domain resource consistency becomes
\[R_{N}\left(\vec{w}\right)=w_{j}R_{N}\left(1,\vec{w}_{-j}\right)+\left(1-w_{j}\right)R_{N}\left(0,\vec{w}_{-j}\right),\quad 0<w_{j}<1.\]

\smallskip

A rule is \emph{efficient} if its shares add up to \(1\), \emph{symmetric} if equal associations receive equal shares, and \emph{null} if it awards nothing to null agents.

\begin{mainthm}[see \cref{thm:shares}]\label{mainthm:shares}
There exists a unique efficient, symmetric, null rule that satisfies resource consistency. For each association vector, all its resource consistency factors coincide and are uniquely determined. The rule extends concede-and-divide.
\end{mainthm}

We call this rule the \emph{maximum-entropy} (\emph{max-entropy}) \emph{rule} \(\operatorname{ME}_{N}\), as we will justify later (\S\ref{sct:me-intro}). For every \(\vec{w}\in\Omega_{N}\) with \(\sum_{i=1}^{N}w_{i}>1\), the proof first shows that the resource consistency factors coincide. Efficiency then determines their common value through the polynomial equation
\[c=1-\prod\nolimits_{i=1}^{N}\left(1-cw_{i}\right)\text{ on }c\in\left(0,1\right].\]
Its unique solution is the \emph{overlap factor} \(c_{\vec{w}}\). The overlap factor equals \(1\) precisely on the total association domain. When \(\sum_{i=1}^{N}w_{i}>1\) and no agent has total association, every resource consistency cut satisfies
\[c_{j}=c_{\vec{w}}<1<c_{j}^{\ast}.\]
The other association proportions are therefore scaled down on the part of \(j\), and scaled up on its complement, where there are fewer agents. When \(\sum_{i=1}^{N}w_{i}=1\), every cut along \(j\) has \(c_{j}=0\).

\subsection{Diskin's principle}

An overlap model is a partition of the resource into regions, each associated with exactly one set of agents, and it is \emph{admissible} if the total proportion of every agent is equal to his association. Diskin's principle treats the associations as \emph{unrelated}: each of them covers the same proportion of every region formed by the others. We call an admissible partition with unrelated associations a \emph{Diskin partition}, after the Talmudist \citet[Vol.~1, \S5]{Diskin1902}, who constructed one for a garment claimed by three people with the proportions \(\left(1,1/2,1/2\right)\) of the example above, and divided each region equally among its claimants, obtaining \(\left(7/12,5/24,5/24\right)\) (\S\ref{sct:diskin-argument}). In probabilistic terms, the associations are unrelated when the events that different agents are associated with a random unit of the resource are independent, as in the \emph{method of random claims} of \citet[\S5.1]{ONeill1982}.

\smallskip

A Diskin partition exists precisely when some agent is associated with the entire resource, and it is then unique (\cref{prop:diskin-partition}). Therefore, on the total association domain, \emph{Diskin's rule} \(\operatorname{Disk}_{N}\) divides every region of the Diskin partition equally among its agents. \Cref{prop:diskin-char} shows that Diskin's rule is the unique efficient, symmetric, null, resource consistent rule on the total association domain, so the rule of \cref{mainthm:shares} extends Diskin's rule. This characterization is the first step in the proof of \cref{mainthm:shares}, since in every resource consistency cut, the cutting agent has total association on his part, so it is governed by Diskin's rule. Off the total association domain, unrelated associations leave an uncovered region of size \(\prod_{i=1}^{N}\left(1-w_{i}\right)\), which rules out Diskin's partition.

\subsection{The max-entropy rule}\label{sct:me-intro}

Identify an overlap model with the distribution \(\mathbb{P}\) of the set \(S\subseteq\left[N\right]\) of agents associated with a random unit of the resource: \(\mathbb{P}\left(S=T\right)\) is the size of the region associated with exactly the members of \(T\). The model is \emph{admissible} for \(\vec{w}\) if
\[\mathbb{P}\left(S\neq\emptyset\right)=1\quad\text{and}\quad\mathbb{P}\left(i\in S\right)=w_{i},\quad i\in\left[N\right].\]
We call such a distribution an \emph{admissible coupling}. The set of all admissible distributions forms a convex polytope we denote by \(\mathcal{P}_{N}\left(\vec{w}\right)\). The independent coupling
\begin{equation}\label{eq:indep}
\mathbb{D}_{\vec{w}}\left(S=T\right)=\Big(\prod\nolimits_{i\in T}w_{i}\Big)\Big(\prod\nolimits_{i\in\left[N\right]\setminus T}\left(1-w_{i}\right)\Big),\quad T\subseteq\left[N\right],
\end{equation}
is admissible precisely on the total association domain, where it describes the Diskin partition.

\smallskip

Maximum (Shannon) entropy is a principle of inference from partial information, famously proposed by \citet[\S2]{jaynes1957}: among the distributions consistent with the data, it selects one of maximal entropy (\S\ref{sct:probmod}), which is seen as the distribution least committed to what the data leave undetermined. \citet{ShoreJohnson1980} derived this principle from consistency requirements on inference. In allocation problems, it was also derived from consistency by \citet{Moulin2016} (\S\ref{sct:Moulin}). In this work, the axioms of \cref{mainthm:shares} characterize the individual shares without specifying an overlap model. We prove that the resulting rule admits a representation by equal division under the unique admissible coupling of maximal entropy.

\begin{mainthm}[see \cref{thm:ent-max,thm:shares}]\label{mainthm:ent-max}
For every \(\vec{w}\in\Omega_{N}\), the unique rule of \cref{mainthm:shares} is given by
\[\operatorname{ME}_{N,i}\left(\vec{w}\right)=\mathbb{E}_{\mathbb{P}_{\vec{w}}^{\ast}}\Big[\mfrac{\mathbf{1}_{\left\{i\in S\right\}}}{\left|S\right|}\Big],\quad i\in\left[N\right],\]
where \(\mathbb{P}_{\vec{w}}^{\ast}\) is the admissible coupling of maximal entropy. When \(\sum_{i=1}^{N}w_{i}>1\), it is given by
\[\mathbb{P}_{\vec{w}}^{\ast}=\mathbb{D}_{c_{\vec{w}}\vec{w}}\left(\cdot\mid S\neq\emptyset\right).\]
\end{mainthm}

The allocation obtained by dividing each unit equally among its associated agents is the Shapley value of the coalitional game induced by the overlap model (\S\ref{sct:alloc}). When \(\sum_{i=1}^{N}w_{i}>1\), the max-entropy rule has the formula
\[\operatorname{ME}_{N,i}\left(\vec{w}\right)=w_{i}\int_{0}^{1}\prod\nolimits_{j\in\left[N\right]\setminus\{i\}}\left(1-c_{\vec{w}}w_{j}+c_{\vec{w}}w_{j}t\right)dt,\quad i\in\left[N\right].\]
Expanding the product gives an integration-free formula (\cref{prop:ent-maxlloc}).

\smallskip

The max-entropy rule is efficient, anonymous, and association bounded. It is cross-association monotone and population monotone in a strict sense (\cref{prop:properties}). It extends concede-and-divide but differs from the Talmud rule. Since the latter is the unique bilaterally consistent extension of concede-and-divide, the max-entropy rule fails bilateral consistency.

\subsection{Related literature}\label{sct:literature}

\subsubsection{Claims problems}

The simple claims problem is due to \citet{ONeill1982}; see \citet{thomson2003,Thomson2015,Thomson2019} for surveys. Besides the Talmud rule \citep{AumannMaschler1985}, extensions of concede-and-divide include the minimal overlap and random arrival rules \citep[\S\S2.1.8--2.1.9]{Thomson2019} and the average-of-awards rule \citep{Miras2022}. A comparison of those rules can be found in \citet{Miras2026}. We compare them and the random stakes rule \citep[\S2.1.10]{Thomson2019} with the max-entropy rule in \S\ref{sct:comparison}. Some works consider randomness of the claim sizes \citep{Xue2018,Long2021}. Our model keeps the claims fixed and places the randomness on their overlaps.

\subsubsection{Museum passes}

For the museum pass, the benchmark is the Shapley rule \citep[\S4]{GinsburghZang2003}, which divides each pass payment equally among the museums visited by its holder. When only the numbers of visitors are known, \citet[\S3]{Bergantinos2015} show that the three axioms characterizing the Shapley rule: symmetry (equal treatment of equals), dummy (nothing to a museum with no visitors), and additivity (with respect to dividing the set of pass holders), are incompatible (Thm.~6). They also characterize the proportional rule (Thm.~7). Other works treat it as a bankruptcy problem \citep{Casas2011,Estevez2012}. Our rule applies the Shapley rule to a reconstruction of the overlaps. It satisfies symmetry and dummy, so it is not additive.

\subsubsection{Subscription platforms}

For subscription platforms, the benchmark is the \emph{user-centric} method with equal splits, which divides the fee of each subscriber equally among his selected providers, and the proportional rule is the \emph{pro-rata} method \citep{Alaei2022,Bergantinos2025}. Both methods are considered by \citet{Bergantinos2026} as rules for claims problems. When only the number of subscribers who selected each provider is available, our rule is the user-centric payment computed on a reconstruction of the overlaps from these aggregates.

\subsubsection{Maximum entropy in allocation}\label{sct:Moulin}

A result conceptually close to ours is that of \citet{Moulin2016}. Students of several types are assigned to schools, and the ideal is zero segregation: all schools have the same composition of types, that is, school and type are independent. Bounds on the number of students of each type in each school can make this proportional assignment infeasible. \citet[Thm.~1(i)]{Moulin2016} shows that among symmetric and continuous assignment rules, consistency (every part of the assignment is chosen in the corresponding smaller problem) together with \emph{constraint neutrality} and the proportional assignment in unconstrained problems, selects the feasible assignment maximizing the entropy
\[-\sum\nolimits_{i,a}y_{ia}\log y_{ia},\]
where \(y_{ia}\) is the number of students of type \(a\) assigned to school \(i\).

\smallskip

In \citet{Moulin2016}, consistency restricts the assignment to fewer types and schools, the proportional assignment is required under no constraints, and entropy is maximized over the feasible assignment matrices. Here, resource consistency cuts the resource and reconstructs the associations on the two parts, and the axioms concern the shares alone without specifying an overlap model. On the total association domain they determine Diskin's rule (\cref{prop:diskin-char}), which is represented by independent associations, and outside this domain independence is incompatible with full coverage. The max-entropy overlap model then gives a representation of the rule by equal division (\cref{mainthm:ent-max}).

\subsection{Organization}

\Cref{sct:diskin} characterizes Diskin's rule on the total association domain. \Cref{sct:probmod,sct:maxent} construct the max-entropy coupling and prove the characterization of the rule. \Cref{sct:claims} returns to contested ownership, discusses the repair of the method of random claims, and compares the max-entropy rule with other claims rules. Appendix~\ref{app:ovfac} establishes the basic properties of the overlap factor and gives the calculation for \cref{lem:factors}; Appendix~\ref{app:entopt} completes the proof of \cref{thm:ent-max}; Appendix~\ref{app:integral} proves the integral identity; Appendix~\ref{app:properties} establishes the properties of the rule; and Appendix~\ref{app:comparison-formulas} derives the formulas for \cref{exm:minimal-overlap} and checks benchmark compatibility. The main notation is summarized in \cref{tab:notation}.

\section{Diskin's rule}\label{sct:diskin}

\subsection{Diskin's argument}\label{sct:diskin-argument}

Diskin \citeyearpar[Vol.~1, \S5]{Diskin1902} considered a garment claimed by three people, the first claiming all of it and each of the other two claiming half of it. In our notation the claims vector is \(\left(1,1/2,1/2\right)\). The division Diskin proposed is
\[\left(14/24,5/24,5/24\right)=\left(7/12,5/24,5/24\right).\]
His argument relies on the principle of the contested garment: what the other claimants concede to a claimant is his without dispute, and what is in dispute is divided equally among its claimants.

\smallskip

\emph{Concessions.} The only concessions are those of the two half-claimants, each concedes the half of the garment he does not claim, and holds that it belongs to the other two claimants, a quarter to each.

\smallskip

\emph{Concession by both.} Both half-claimants concede a quarter to the total-claimant, and neither concedes him more. Thus the two half-claimants agree that he is entitled to a quarter, and so he takes it.

\smallskip

\emph{Concession by one.} The third claimant concedes a quarter to the second, but the total-claimant does not concede it, and so the two divide it equally, \(1/8\) each. In the same way, the first and the third claimants divide the quarter which the second concedes to the third. At this stage the total-claimant holds \(1/4+1/8+1/8=1/2\) and each half-claimant holds \(1/8\).

\smallskip

\emph{Concession by none.} The remaining quarter is conceded by no one and claimed by all three, so it is divided equally, \(1/12\) each.

\smallskip

The argument contains a single modeling step: a half-claimant who concedes half cannot say which half the other half-claimant means, and he takes the other's claim to cover half of what he concedes and half of what he claims. The two partial claims are then treated as unrelated, which is Diskin's principle. The garment then falls into four equal regions, claimed respectively by \(\left\{1\right\}\), \(\left\{1,2\right\}\), \(\left\{1,3\right\}\), \(\left\{1,2,3\right\}\), which is the Diskin partition of \(\left(1,1/2,1/2\right)\) (\S\ref{sct:diskin-partitions}), and the three layers are the equal division of each region among its claimants. The total-claimant guarantees that no region is conceded by everyone.

\subsection{Diskin partitions}\label{sct:diskin-partitions}

Diskin's argument applies to \cref{prob:overlaps} whenever some agent is associated with the entire resource, and we now show that this is exactly when it applies. An association vector is any \(\vec{w}\in\left[0,1\right]^{N}\). The \emph{total association domain} of the allocation problem is
\[\Omega_{N}^{\mathrm{tot}}\coloneq\big\{\vec{w}\in\Omega_{N}:\max\nolimits_{i\in\left[N\right]}w_{i}=1\big\}.\]

A \emph{partition} of the resource is described by \(\left(p_{T}\right)_{T\subseteq\left[N\right]}\), where \(0\leq p_{T}\leq 1\) is the size of the region associated with exactly the members of \(T\), such that \(p_{\emptyset}=0\) (every unit is associated) and \(\sum_{T\subseteq\left[N\right]}p_{T}=1\). It is \emph{admissible} for \(\vec{w}\) if the total proportion of the association of every agent is his original association:
\[\sum\nolimits_{\left\{T\subseteq\left[N\right]:i\in T\right\}}p_{T}=w_{i},\quad i\in\left[N\right].\]

\smallskip

Equivalently, if \(S\) is the set of agents associated with a random unit of the resource, then \(p_{T}=\mathbb{P}\left(S=T\right)\). We use the partition and its probability distribution as two descriptions of the same overlap model. Its association indicators are \(X_{i}\coloneq\mathbf{1}_{\left\{i\in S\right\}}\), \(i\in\left[N\right]\).

\begin{defn}\label{dfn:diskin-partition}
An admissible partition is a \emph{Diskin partition} if adding the same agent to two sets of agents preserves the relative sizes of their regions. That is, for every \(j\in\left[N\right]\) and every \(T_{1},T_{2}\subseteq\left[N\right]\setminus\{j\}\),
\[p_{T_{1}\cup\left\{j\right\}}p_{T_{2}}=p_{T_{1}}p_{T_{2}\cup\left\{j\right\}}.\]
\end{defn}

\begin{prop}\label{prop:diskin-partition}
An association vector \(\vec{w}\in\left[0,1\right]^{N}\) has a Diskin partition if and only if \(\max_{i\in\left[N\right]}w_{i}=1\). In this case the Diskin partition is unique, and its region sizes are given by the independent coupling \eqref{eq:indep}:
\begin{equation}\label{eq:diskin-regions}
p_{T}^{\vec{w}}\coloneq\mathbb{D}_{\vec{w}}\left(S=T\right),\quad T\subseteq\left[N\right].
\end{equation}
\end{prop}

We use the weights \(p_{T}^{\vec{w}}\) in \eqref{eq:diskin-regions} for every \(\vec{w}\in\left[0,1\right]^{N}\), whether or not they form an admissible partition.

\begin{proof}[Proof of \cref{prop:diskin-partition}]
We shall think of this in a probabilistic way. Consider the association indicators \(X_{i}=\mathbf{1}_{\left\{i\in S\right\}}\), which record whether the \(i^{\mathrm{th}}\) agent is associated with a random unit of the resource. An admissible partition \(\left(p_{T}\right)_{T\subseteq\left[N\right]}\) induces a joint distribution \(\mathbb{P}\) of the random vector \(\left(X_{1},\dotsc,X_{N}\right)\) via
\[\mathbb{P}\left(X_{i}=1\text{ exactly when }i\in T\right)\coloneq p_{T},\quad T\subseteq\left[N\right].\]
By admissibility one has
\[\mathbb{P}\left(X_{i}=1\right)=\sum\nolimits_{\left\{T\subseteq\left[N\right]:i\in T\right\}}p_{T}=w_{i},\quad i\in\left[N\right],\]
and so \(X_{i}\) is a Bernoulli random variable with probability \(w_{i}\) of success.

\smallskip

Suppose now that \(\left(p_{T}\right)_{T\subseteq\left[N\right]}\) is a Diskin partition. Fix \(j\in\left[N\right]\). For every \(T\subseteq\left[N\right]\setminus\{j\}\), summing the defining identity of Diskin partition with \(T_{1}=T\) over \(T_{2}\subseteq\left[N\right]\setminus\left\{j\right\}\) gives
\[p_{T\cup\left\{j\right\}}\left(1-w_{j}\right)=p_{T}w_{j}.\]
If \(w_{j}\in\left\{0,1\right\}\), then \(X_{j}\) is constant and hence independent of the other indicators. If \(0<w_{j}<1\), dividing by \(1-w_{j}\) and by \(w_{j}\), this identity becomes
\[\mathbb{P}\left(S\setminus\left\{j\right\}=T\mid X_{j}=1\right)=\mathbb{P}\left(S\setminus\left\{j\right\}=T\mid X_{j}=0\right).\]
Thus the indicators are mutually independent with marginals \(\vec{w}\), so \(p_{T}=\mathbb{D}_{\vec{w}}\left(S=T\right)=p_{T}^{\vec{w}}\) for every \(T\subseteq\left[N\right]\). This proves the uniqueness. Now for every \(\vec{w}\in\left[0,1\right]^{N}\), one sees that \(\left(p_{T}^{\vec{w}}\right)_{T\subseteq\left[N\right]}\) satisfy the defining identity of Diskin partition. Since they are the joint probabilities of independent Bernoulli indicators with probabilities of success \(w_{i}\), they sum to \(1\) and give each agent his prescribed association:
\begin{equation}\label{eq:diskin-marginals}
\sum\nolimits_{T\subseteq\left[N\right]}p_{T}^{\vec{w}}=1\quad\text{and}\quad\sum\nolimits_{\left\{T\subseteq\left[N\right]:i\in T\right\}}p_{T}^{\vec{w}}=w_{i}.
\end{equation}
Finally, we have \(p_{\emptyset}^{\vec{w}}=\prod\nolimits_{i=1}^{N}\left(1-w_{i}\right)\), and thus \(p_{\emptyset}^{\vec{w}}=0\) precisely when \(\max_{i\in\left[N\right]}w_{i}=1\).
\end{proof}

\subsection{Diskin's rule}\label{sct:diskin-rule}

\begin{defn}\label{dfn:diskin-rule}
Diskin's rule \(\left(\operatorname{Disk}_{N,i}\right)_{N\geq 1}\) is defined on \(\Omega_{N}^{\mathrm{tot}}\) by
\[\operatorname{Disk}_{N,i}\left(\vec{w}\right)\coloneq\sum\nolimits_{\left\{T\subseteq\left[N\right]:i\in T\right\}}\mfrac{p_{T}^{\vec{w}}}{\left|T\right|},\quad i\in\left[N\right].\]
\end{defn}

The following proposition is an important intermediate step in proving \cref{mainthm:shares}. Let us establish some standard terminology (see \citet[\S3.1, \S3.4.1, \S3.6]{Thomson2019}). A rule \(\left(R_{N}\right)_{N\geq 1}\) is called:
\begin{itemize}
    \item \emph{Efficient} if \(\sum_{i=1}^{N}R_{N,i}\left(\vec{w}\right)=1\).
    \item \emph{Anonymous} if \(R_{N,\sigma\left(i\right)}\left(\sigma\vec{w}\right)=R_{N,i}\left(\vec{w}\right)\) for every permutation \(\sigma\) of \(\left[N\right]\).
    \item \emph{Symmetric} if \(R_{N,i}\left(\vec{w}\right)=R_{N,j}\left(\vec{w}\right)\) whenever \(w_{i}=w_{j}\) (a weaker form of anonymity).
\end{itemize}

\begin{prop}\label{prop:diskin-char}
Diskin's rule is the unique efficient, symmetric, null rule on \(\Omega_{N}^{\mathrm{tot}}\) that satisfies resource consistency. For every \(\vec{w}\in\Omega_{N}^{\mathrm{tot}}\) and \(i\in\left[N\right]\) it is given by
\begin{equation}\label{eq:diskin-integral}
\operatorname{Disk}_{N,i}\left(\vec{w}\right)=w_{i}\int_{0}^{1}\prod\nolimits_{j\in\left[N\right]\setminus\{i\}}\left(1-w_{j}+w_{j}t\right)dt=w_{i}\sum\nolimits_{J\subseteq\left[N\right]\setminus\left\{i\right\}}\mfrac{\left(-1\right)^{\left|J\right|}}{\left|J\right|+1}\prod\nolimits_{j\in J}w_{j}.
\end{equation}
\end{prop}

\begin{proof}
Note that \(p_{\sigma\left(T\right)}^{\sigma\vec{w}}=p_{T}^{\vec{w}}\), so Diskin's rule is anonymous, hence symmetric. It is efficient by \eqref{eq:diskin-marginals}, and null by definition. For resource consistency, fix \(\vec{w}\in\Omega_{N}^{\mathrm{tot}}\) and \(j\in\left[N\right]\) with \(0<w_{j}<1\), and take \(c_{j}=1\), so that \(c_{j}^{\ast}=1\). Note that \(\left(1,\vec{w}_{-j}\right),\left(0,\vec{w}_{-j}\right)\in\Omega_{N}^{\mathrm{tot}}\), so by isolating the factor for \(j\) in \eqref{eq:indep} and using \eqref{eq:diskin-regions} we get
\[p_{T}^{\vec{w}}=w_{j}\,p_{T}^{\left(1,\vec{w}_{-j}\right)}+\left(1-w_{j}\right)p_{T}^{\left(0,\vec{w}_{-j}\right)},\quad T\subseteq\left[N\right].\]
This readily implies that
\[\operatorname{Disk}_{N}\left(\vec{w}\right)=w_{j}\operatorname{Disk}_{N}\left(1,\vec{w}_{-j}\right)+\left(1-w_{j}\right)\operatorname{Disk}_{N}\left(0,\vec{w}_{-j}\right).\]

\smallskip

Conversely, let \(\left(R_{N}\right)_{N\geq1}\) satisfy the four requirements on \(\Omega_{N}^{\mathrm{tot}}\), and fix \(N\geq1\). We prove \(R_{N}\left(\vec{w}\right)=\operatorname{Disk}_{N}\left(\vec{w}\right)\) by induction on
\[m\left(\vec{w}\right)\coloneq\#\{i\in\left[N\right]:0<w_{i}<1\}.\]
If \(m\left(\vec{w}\right)=0\), all coordinates belong to \(\left\{0,1\right\}\). Nullity gives zero to the null agents, and symmetry and efficiency divide the resource equally among the agents with total association. This is the same as Diskin's rule. Assume then that \(m\left(\vec{w}\right)\geq1\), and fix \(j\) with \(0<w_{j}<1\). Admissibility gives \(c_{j}=c_{j}^{\ast}=1\), as shown after \cref{dfn:resource-consistency}. Both \(\left(1,\vec{w}_{-j}\right)\) and \(\left(0,\vec{w}_{-j}\right)\) belong to \(\Omega_{N}^{\mathrm{tot}}\), both have \(m\left(\vec{w}\right)-1\) coordinates in \(\left(0,1\right)\). Using resource consistency and the induction hypothesis we get
\begin{align*}
R_{N}\left(\vec{w}\right)
&=w_{j}R_{N}\left(1,\vec{w}_{-j}\right)+\left(1-w_{j}\right)R_{N}\left(0,\vec{w}_{-j}\right)\\
&=w_{j}\operatorname{Disk}_{N}\left(1,\vec{w}_{-j}\right)+\left(1-w_{j}\right)\operatorname{Disk}_{N}\left(0,\vec{w}_{-j}\right)
=\operatorname{Disk}_{N}\left(\vec{w}\right),
\end{align*}
where the last equality is by the resource consistency of Diskin's rule.

\smallskip

Finally, the formula \eqref{eq:diskin-integral} follows from \cref{dfn:diskin-rule} through the identity \eqref{eq:integral} with \(\vec{r}=\vec{w}\).
\end{proof}

\section{Overlap models and maximum entropy}\label{sct:probmod}

Diskin's rule represents the shares determined by resource consistency on the total association domain. Its independent overlap model is not admissible outside this domain. We now characterize the max-entropy admissible coupling, before returning to the shares it induces.

\smallskip

Fix \(N\geq1\). We use the probabilistic description of overlap models introduced in \S\ref{sct:diskin-partitions}, and write
\[\mathcal{S}_{N}\coloneq2^{\left[N\right]}\setminus\left\{\emptyset\right\},\quad\mathbb{P}\left(T\right)\coloneq\mathbb{P}\left(S=T\right).\]
Let \(\operatorname{Prob}\left(\mathcal{S}_{N}\right)\) be the set of distributions on \(\mathcal{S}_{N}\). For \(\vec{w}\in\Omega_{N}\), the set of admissible couplings is
\[\mathcal{P}_{N}\left(\vec{w}\right)\coloneq\left\{\mathbb{P}\in\operatorname{Prob}\left(\mathcal{S}_{N}\right):\mathbb{P}\left(i\in S\right)=w_{i}\text{ for every }i\in\left[N\right]\right\}.\]
The exclusion of the empty set expresses full coverage. By \eqref{eq:indep}, the independent coupling \(\mathbb{D}_{\vec{w}}\) is admissible precisely when \(\max_{i\in\left[N\right]}w_{i}=1\). On this domain we have \(\mathbb{D}_{\vec{w}}\left(T\right)=p_{T}^{\vec{w}}\), so Diskin's rule is equal division under the independent coupling:
\begin{equation}\label{eq:diskin-expectation}
\operatorname{Disk}_{N,i}\left(\vec{w}\right)=\mathbb{E}_{\mathbb{D}_{\vec{w}}}\Big[\mfrac{\mathbf{1}_{\left\{i\in S\right\}}}{\left|S\right|}\Big],\quad i\in\left[N\right].
\end{equation}

\smallskip

The (Shannon) \emph{entropy} of a probability distribution \(\mathbb{Q}\) on a finite set \(\mathcal{S}\) is the quantity
\[H\left(\mathbb{Q}\right)\coloneq-\sum\nolimits_{s\in\mathcal{S}}\mathbb{Q}\left(s\right)\log\mathbb{Q}\left(s\right),\quad(0\log0\coloneq0).\]

\smallskip

For \(\vec{w}\in\Omega_{N}\), consider the optimization problem
\[\mathbb{P}_{\vec{w}}^{\ast}=\arg\max\left\{H\left(\mathbb{P}\right):\mathbb{P}\in\mathcal{P}_{N}\left(\vec{w}\right)\right\}.\]
This is a max-entropy problem with prescribed marginals over \(\mathcal{S}_{N}\) \citep[\S3]{SinghVishnoi2014}. \Cref{thm:ent-max} identifies its solution using a single scalar, the overlap factor.

\smallskip

Regard every \(\mathbb{P}\in\mathcal{P}_{N}\left(\vec{w}\right)\) as a distribution on \(2^{\left[N\right]}\) by setting \(\mathbb{P}\left(\emptyset\right)=0\). The marginal constraints give \(\mathbb{P}\left(T\right)=0\) whenever \(\mathbb{D}_{\vec{w}}\left(T\right)=0\), so the relative entropy of \(\mathbb{P}\) with respect to \(\mathbb{D}_{\vec{w}}\) is given by
\[\operatorname{D}_{\mathrm{KL}}\left(\mathbb{P}\mathbin{\Vert}\mathbb{D}_{\vec{w}}\right)\coloneq\sum\nolimits_{\left\{T\in\mathcal{S}_{N}:\mathbb{P}\left(T\right)>0\right\}}\mathbb{P}\left(T\right)\log\mfrac{\mathbb{P}\left(T\right)}{\mathbb{D}_{\vec{w}}\left(T\right)}.\]
It is also called the \emph{Kullback--Leibler divergence}, and here it is the \emph{total correlation} of the association indicators \citep[\S2]{Watanabe1960}. As observed by Watanabe there, since the marginals are fixed,
\begin{equation}\label{eq:total-correlation}
\operatorname{D}_{\mathrm{KL}}\left(\mathbb{P}\mathbin{\Vert}\mathbb{D}_{\vec{w}}\right)=H\left(\mathbb{D}_{\vec{w}}\right)-H\left(\mathbb{P}\right),\quad\mathbb{P}\in\mathcal{P}_{N}\left(\vec{w}\right).
\end{equation}

\smallskip

For \(\vec{w}\in\Omega_{N}\) with \(\sum_{i=1}^{N}w_{i}>1\), the max-entropy coupling will be obtained by conditioning a certain product distribution on full coverage. The prescribed marginals determine its parameters up to a single scalar. Indeed, if the product parameters are \(\vec{r}\in\left[0,1\right]^{N}\) and \(c\coloneq\mathbb{D}_{\vec{r}}\left(S\neq\emptyset\right)>0\), admissibility requires
\[w_{i}=\mathbb{D}_{\vec{r}}\left(i\in S\mid S\neq\emptyset\right)=\mfrac{r_{i}}{c},\quad i\in\left[N\right].\]
Thus \(\vec{r}=c\vec{w}\), and the identity \(c=\mathbb{D}_{c\vec{w}}\left(S\neq\emptyset\right)\) gives the polynomial equation below. The same equation will arise from efficiency in \cref{lem:factors}, once the resource consistency factors are shown to coincide. Its existence, uniqueness, and basic properties are established in Appendix~\ref{app:ovfac}.

\begin{defn}\label{dfn:overfac}
For \(\vec{w}\in\Omega_{N}\) with \(\sum_{i=1}^{N}w_{i}>1\), the \emph{overlap factor} \(c_{\vec{w}}\) is the unique solution in \(\left(0,1\right]\) of the polynomial equation
\[c=1-\prod\nolimits_{i=1}^{N}\left(1-cw_{i}\right).\]
The overlap factor satisfies \(c_{\vec{w}}=1\) if and only if \(\max_{i\in\left[N\right]}w_{i}=1\) (Appendix~\ref{app:ovfac}).
\end{defn}

The following theorem characterizes the max-entropy coupling.

\begin{thm}\label{thm:ent-max}
For every \(\vec{w}\in\Omega_{N}\), the polytope \(\mathcal{P}_{N}\left(\vec{w}\right)\) contains a unique max-entropy element \(\mathbb{P}_{\vec{w}}^{\ast}\). When \(\sum_{i=1}^{N}w_{i}=1\), the admissible coupling is unique and is given by
\[\mathbb{P}_{\vec{w}}^{\ast}\left(\left\{i\right\}\right)=w_{i},\quad i\in\left[N\right].\]
When \(\sum_{i=1}^{N}w_{i}>1\), the following conditions are equivalent, and each characterizes \(\mathbb{P}_{\vec{w}}^{\ast}\) uniquely:
\begin{enumerate}
    \item \(\mathbb{P}\) maximizes entropy \(H\left(\mathbb{P}\right)\) on \(\mathcal{P}_{N}\left(\vec{w}\right)\).
    \item \(\mathbb{P}\left(T\right)>0\) for every \(T\in\mathcal{S}_{N}\) with \(\mathbb{D}_{\vec{w}}\left(T\right)>0\),\footnote{Note that \(\mathbb{D}_{\vec{w}}\left(T\right)>0\) precisely when \(T\) contains every \(i\) with \(w_{i}=1\) and excludes every \(j\) with \(w_{j}=0\).} and for every \(i\in\left[N\right]\) with \(w_{i}>0\), the indicators
    \[X_{j}\mid\{X_{i}=1\},\quad j\in\left[N\right]\setminus\{i\},\]
    are mutually independent under \(\mathbb{P}\).\footnote{This means that the association indicators \(X_{j}\), \(j\in\left[N\right]\setminus\{i\}\), are mutually independent under \(\mathbb{P}\left(\cdot\mid X_{i}=1\right)\).}
    \item \(\mathbb{P}\) is the distribution
    \[\mathbb{P}\left(T\right)=\mathbb{D}_{c_{\vec{w}}\vec{w}}\left(T\mid S\neq\emptyset\right),\]
    where \(c_{\vec{w}}\) is the overlap factor as in \cref{dfn:overfac} and \(c_{\vec{w}}\vec{w}=\left(c_{\vec{w}}w_{1},\dotsc,c_{\vec{w}}w_{N}\right)\). More explicitly,
    \begin{equation}\label{eq:maxent-coupling}
    \mathbb{P}\left(T\right)=\mfrac{1}{c_{\vec{w}}}\Big(\prod\nolimits_{i\in T}c_{\vec{w}}w_{i}\Big)\Big(\prod\nolimits_{i\in\left[N\right]\setminus T}\left(1-c_{\vec{w}}w_{i}\right)\Big),\quad T\in\mathcal{S}_{N}.
    \end{equation}
\end{enumerate}
On the total association domain \(\mathbb{P}_{\vec{w}}^{\ast}=\mathbb{D}_{\vec{w}}\), and the max-entropy coupling is the Diskin partition.
\end{thm}

\begin{proof}
If \(\sum_{i=1}^{N}w_{i}=1\), every admissible coupling satisfies \(\mathbb{E}_{\mathbb{P}}\left[\left|S\right|\right]=1\), hence \(\left|S\right|=1\) almost surely, and the marginal constraints determine the coupling. Assume then that \(\sum_{i=1}^{N}w_{i}>1\). An agent with zero association can be ignored under every admissible coupling. Deleting all such agents preserves entropy and each condition of the theorem. We then assume that \(w_{i}>0\) for every \(i\).

\smallskip

Suppose first that \(\max_{i\in\left[N\right]}w_{i}=1\). Then \(c_{\vec{w}}=1\) and \(\mathbb{D}_{\vec{w}}\in\mathcal{P}_{N}\left(\vec{w}\right)\). By \eqref{eq:total-correlation} and Gibbs' inequality \citep[Thm.~2.6.3]{CoverThomas2006}, \(\mathbb{D}_{\vec{w}}\) is the unique entropy maximizer. It satisfies (2), and since \(\mathbb{D}_{\vec{w}}\left(S\neq\emptyset\right)=1\) it is the distribution in (3). Conversely, if \(\mathbb{P}\in\mathcal{P}_{N}\left(\vec{w}\right)\) satisfies (2), take any \(k\) with \(w_{k}=1\), so \(X_{k}=1\) almost surely and conditioning \(\mathbb{P}\) on \(X_{k}=1\) does not change \(\mathbb{P}\). Thus (2) gives mutual independence of the other indicators. Together with the constant indicator \(X_{k}\), this gives \(\mathbb{P}=\mathbb{D}_{\vec{w}}\).

\smallskip

It remains to consider the case \(\max_{i\in\left[N\right]}w_{i}<1\). We first construct a conditioned product distribution with the prescribed marginals, and then prove that it maximizes entropy. Put
\[c\coloneq c_{\vec{w}}\quad\text{and}\quad\vec{r}\coloneq c\vec{w}.\]
By the defining polynomial equation of the overlap factor,
\[\mathbb{D}_{\vec{r}}\left(S\neq\emptyset\right)=1-\prod\nolimits_{i=1}^{N}\left(1-r_{i}\right)=c.\]
Let \(\mathbb{Q}\left(\cdot\right)\coloneq\mathbb{D}_{\vec{r}}\left(\cdot\mid S\neq\emptyset\right)\) be the distribution in \eqref{eq:maxent-coupling}. Since \(i\in S\) implies \(S\neq\emptyset\), its marginals are
\[\mathbb{Q}\left(X_{i}=1\right)=\mathbb{Q}\left(i\in S\right)=\mfrac{\mathbb{D}_{\vec{r}}\left(i\in S\right)}{\mathbb{D}_{\vec{r}}\left(S\neq\emptyset\right)}=\mfrac{r_{i}}{c}=w_{i},\quad i\in\left[N\right].\]
Thus \(\mathbb{Q}\in\mathcal{P}_{N}\left(\vec{w}\right)\). To show that \(\mathbb{Q}\) maximizes entropy, we compare it with an arbitrary admissible coupling. Since \(0<r_{i}<1\) for every \(i\), \(\mathbb{Q}\) has full support on \(\mathcal{S}_{N}\), and
\[\log\mathbb{Q}\left(S\right)=-\log c+\sum\nolimits_{i=1}^{N}\left(X_{i}\log r_{i}+\left(1-X_{i}\right)\log\left(1-r_{i}\right)\right).\]
For every \(\mathbb{P}\in\mathcal{P}_{N}\left(\vec{w}\right)\), admissibility gives \(\mathbb{E}_{\mathbb{P}}\left[X_{i}\right]=\mathbb{E}_{\mathbb{Q}}\left[X_{i}\right]=w_{i}\) for every \(i\). Then similarly to \eqref{eq:total-correlation},
\[\operatorname{D}_{\mathrm{KL}}\left(\mathbb{P}\mathbin{\Vert}\mathbb{Q}\right)=-H\left(\mathbb{P}\right)-\mathbb{E}_{\mathbb{P}}\left[\log\mathbb{Q}\left(S\right)\right]=-H\left(\mathbb{P}\right)-\mathbb{E}_{\mathbb{Q}}\left[\log\mathbb{Q}\left(S\right)\right]=H\left(\mathbb{Q}\right)-H\left(\mathbb{P}\right).\]
By Gibbs' inequality, it is nonnegative with equality only when \(\mathbb{P}=\mathbb{Q}\), so \(\mathbb{Q}\) uniquely maximizes entropy.

\smallskip

We next verify (2). Fix \(i\in\left[N\right]\). Since \(X_{i}=1\) implies \(S\neq\emptyset\),
\[\mathbb{Q}\left(\cdot\mid X_{i}=1\right)=\mathbb{D}_{\vec{r}}\left(\cdot\mid X_{i}=1\right).\]
Under the distribution \(\mathbb{D}_{\vec{r}}\), conditioning on \(X_{i}=1\) makes the other indicators mutually independent, with probabilities of success \(\vec{r}_{-i}=\left(r_{j}\right)_{j\in\left[N\right]\setminus\{i\}}\). Together with the full support of \(\mathbb{Q}\), this proves (2).

\smallskip

The implication \((2)\Rightarrow(3)\) in the case \(\max_{i\in\left[N\right]}w_{i}<1\) is proved in Appendix~\ref{app:entopt}.
\end{proof}

Let us elaborate on the meanings of the different characterizations in \cref{thm:ent-max}.
\begin{itemize}
    \item Characterization (1) has an information-theoretic meaning. By \eqref{eq:total-correlation}, maximizing entropy is equivalent to minimizing total correlation, so the selected coupling is closest to independence in relative entropy, the I-projection of the independent coupling \citep[\S\S2--3]{Csiszar1975}.
    \item Characterization (2) expresses Diskin's principle within the part associated with each agent: the other associations are independent on that part. By (3), their conditional proportions are
    \[\mathbb{P}_{\vec{w}}^{\ast}\left(k\in S\mid j\in S\right)=c_{\vec{w}}w_{k},\quad k\neq j.\]
    Thus the part of \(j\) carries the Diskin partition of \(\left(1,c_{\vec{w}}\vec{w}_{-j}\right)\), but independence is not required on the complementary part. Nevertheless, that complementary part carries the max-entropy coupling of its reconstructed associations (\cref{prop:split}).
    \item Characterization (3) relates the max-entropy coupling to the method of random claims. As shown in \S\ref{sct:randomclaims}, applying that method to the rescaled proportions \(c_{\vec{w}}\vec{w}\) gives the max-entropy coupling with the original proportions \(\vec{w}\).
\end{itemize}

The following proposition shows that the max-entropy coupling is consistent with cuts, which will lead to the resource consistency of the max-entropy rule.

\begin{prop}\label{prop:split}
Let \(\vec{w}\in\Omega_{N}\) with \(\sum_{i=1}^{N}w_{i}>1\), put \(c\coloneq c_{\vec{w}}\), and let \(j\in\left[N\right]\) with \(0<w_{j}<1\). Then
\[\mathbb{P}_{\vec{w}}^{\ast}\left(\cdot\mid j\in S\right)=\mathbb{P}_{\left(1,c\vec{w}_{-j}\right)}^{\ast}=\mathbb{D}_{\left(1,c\vec{w}_{-j}\right)}\quad\text{and}\quad\mathbb{P}_{\vec{w}}^{\ast}\left(\cdot\mid j\notin S\right)=\mathbb{P}_{\left(0,c_{j}^{\ast}\vec{w}_{-j}\right)}^{\ast},\]
where \(c_{j}^{\ast}=\left(1-cw_{j}\right)/\left(1-w_{j}\right)\) \eqref{eq:cjast}. In particular, for every \(k\neq j\),
\[\mathbb{P}_{\vec{w}}^{\ast}\left(k\in S\mid j\in S\right)=cw_{k}\quad\text{and}\quad\mathbb{P}_{\vec{w}}^{\ast}\left(k\in S\mid j\notin S\right)=c_{j}^{\ast}w_{k}.\]
\end{prop}

\begin{proof}
By \cref{thm:ent-max}(3) we have \(\mathbb{P}_{\vec{w}}^{\ast}=\mathbb{D}_{c\vec{w}}\left(\cdot\mid S\neq\emptyset\right)\). Since \(X_{j}=1\) implies \(S\neq\emptyset\), we get
\[\mathbb{P}_{\vec{w}}^{\ast}\left(\cdot\mid X_{j}=1\right)=\mathbb{D}_{c\vec{w}}\left(\cdot\mid X_{j}=1\right)=\mathbb{D}_{\left(1,c\vec{w}_{-j}\right)}=\mathbb{P}_{\left(1,c\vec{w}_{-j}\right)}^{\ast}.\]
Then for \(k\neq j\) we have \(\mathbb{P}_{\vec{w}}^{\ast}\left(X_{k}=1\mid X_{j}=1\right)=cw_{k}w_{j}/w_{j}=cw_{k}\). Conditioning on \(X_{j}=0\) gives
\[\mathbb{P}_{\vec{w}}^{\ast}\left(\cdot\mid X_{j}=0\right)=\mathbb{D}_{c\vec{w}}\left(\cdot\mid X_{j}=0,S\neq\emptyset\right)=\mathbb{D}_{\left(0,c\vec{w}_{-j}\right)}\left(\cdot\mid S\neq\emptyset\right).\]
Using the definition of the overlap factor \(c\), the probability of the conditioning event \(\left\{S\neq\emptyset\right\}\) is
\[d\coloneq\mathbb{D}_{\left(0,c\vec{w}_{-j}\right)}\left(S\neq\emptyset\right)=1-\prod\nolimits_{k\in\left[N\right]\setminus\{j\}}\left(1-cw_{k}\right)=1-\mfrac{1-c}{1-cw_{j}}=\mfrac{c}{c_{j}^{\ast}}.\]
Then for \(k\neq j\) we have
\[\mathbb{P}_{\vec{w}}^{\ast}\left(X_{k}=1\mid X_{j}=0\right)=\mfrac{\mathbb{D}_{\left(0,c\vec{w}_{-j}\right)}\left(X_{k}=1\right)}{d}=\mfrac{cw_{k}}{d}=c_{j}^{\ast}w_{k}.\]

\smallskip

The first identity in the proposition is then established. For the second identity, put \(\vec{u}\coloneq\left(0,c_{j}^{\ast}\vec{w}_{-j}\right)\). The preceding calculation shows that \(\mathbb{P}_{\vec{w}}^{\ast}\left(\cdot\mid X_{j}=0\right)\) is an admissible coupling of \(\vec{u}\), and it remains to show that it is \(\mathbb{P}_{\vec{u}}^{\ast}\). When \(\sum_{i=1}^{N}u_{i}=1\) there is a unique admissible coupling. Otherwise, since \(d\vec{u}=\left(0,c\vec{w}_{-j}\right)\) we have
\[d=\mathbb{D}_{d\vec{u}}\left(S\neq\emptyset\right)=1-\prod\nolimits_{i=1}^{N}\left(1-du_{i}\right).\]
Since \(d\in\left(0,1\right]\), \cref{dfn:overfac} gives \(d=c_{\vec{u}}\). By \cref{thm:ent-max}(3),
\[\mathbb{P}_{\vec{w}}^{\ast}\left(\cdot\mid X_{j}=0\right)=\mathbb{D}_{c_{\vec{u}}\vec{u}}\left(\cdot\mid S\neq\emptyset\right)=\mathbb{P}_{\vec{u}}^{\ast}.\qedhere\]
\end{proof}

\section{The max-entropy rule}\label{sct:maxent}

\subsection{The full-information benchmark}\label{sct:alloc}

For each \(N\geq 1\), define the allocation map
\[\Phi_{N}:\operatorname{Prob}\left(\mathcal{S}_{N}\right)\longrightarrow\left[0,1\right]^{N}\]
by dividing each unit equally among its associated agents. It is given by the formula
\begin{equation}\label{eq:alloceuiv}
\Phi_{N,i}\left(\mathbb{P}\right)=\mathbb{E}_{\mathbb{P}}\left[\mfrac{\mathbf{1}_{\left\{i\in S\right\}}}{\left|S\right|}\right]=\sum\nolimits_{\left\{T\in\mathcal{S}_{N}:i\in T\right\}}\mfrac{\mathbb{P}\left(T\right)}{\left|T\right|},\quad i\in\left[N\right].
\end{equation}

\smallskip

This allocation can be recast as the Shapley value of an appropriate coalitional game. Every distribution \(\mathbb{P}\in\operatorname{Prob}\left(\mathcal{S}_{N}\right)\) induces the coalitional game
\[v_{\mathbb{P}}:2^{\left[N\right]}\longrightarrow\mathbb{R},\quad v_{\mathbb{P}}\left(A\right)\coloneq\mathbb{P}\left(S\subseteq A\right).\]
Thus \(v_{\mathbb{P}}\left(A\right)\) is the proportion of the resource with which no agent outside \(A\) is associated. In \cref{prob:Ncontgarm}, it is the part conceded to \(A\) by the other claimants. Note that we can write
\[v_{\mathbb{P}}=\sum\nolimits_{T\in\mathcal{S}_{N}}\mathbb{P}\left(T\right)u_{T},\quad u_{T}\left(A\right)\coloneq\mathbf{1}_{\left\{T\subseteq A\right\}}.\]
The Shapley value of the game \(u_{T}\) assigns \(1/\left|T\right|\) to each member of \(T\) and zero to all others, so the representation follows by the linearity of the Shapley value; see \citet[\S5.5]{Moulin2003}.

\smallskip

For \(\mathbb{P}\in\operatorname{Prob}\left(\mathcal{S}_{N}\right)\) and \(j\in\left[N\right]\), put \(w_{j}\coloneq\mathbb{P}\left(j\in S\right)\) and assume that \(0<w_{j}<1\). By the law of total expectation, for every \(i\in\left[N\right]\) we have
\begin{align*}
\Phi_{N,i}\left(\mathbb{P}\right)
&=w_{j}\mathbb{E}_{\mathbb{P}}\Big[\frac{\mathbf{1}_{\left\{i\in S\right\}}}{\left|S\right|}\mid j\in S\Big]+\left(1-w_{j}\right)\mathbb{E}_{\mathbb{P}}\Big[\frac{\mathbf{1}_{\left\{i\in S\right\}}}{\left|S\right|}\mid j\notin S\Big]\\
&=w_{j}\Phi_{N,i}\left(\mathbb{P}\left(\cdot\mid j\in S\right)\right)
+\left(1-w_{j}\right)\Phi_{N,i}\left(\mathbb{P}\left(\cdot\mid j\notin S\right)\right).
\end{align*}
Therefore, the allocation \(\left(\Phi_{N}\right)_{N\geq 1}\) respects cuts of the resource:
\begin{equation}\label{eq:benchmark-cut}
\Phi_{N}\left(\mathbb{P}\right)=w_{j}\Phi_{N}\left(\mathbb{P}\left(\cdot\mid j\in S\right)\right)+\left(1-w_{j}\right)\Phi_{N}\left(\mathbb{P}\left(\cdot\mid j\notin S\right)\right).
\end{equation}

\smallskip

From the formulation of \cref{prob:overlaps}, one expects that a rule for it would be compatible with our full-information benchmark, in the sense that it would be equal division of each unit under some admissible overlap model. Let us define this precisely.

\begin{defn}\label{dfn:bench-comp}
A rule \(\left(R_{N}\right)_{N\geq1}\) is said to be compatible with our \emph{full-information benchmark} if for every \(\vec{w}\in\Omega_{N}\) there exists some admissible coupling \(\mathbb{P}\in\mathcal{P}_{N}\left(\vec{w}\right)\) such that
\[R_{N}\left(\vec{w}\right)=\Phi_{N}\left(\mathbb{P}\right),\]
where \(\Phi_{N}\) is given in \eqref{eq:alloceuiv}.
\end{defn}

\subsection{The rule}\label{sct:maxent-rule}

We can now combine the max-entropy admissible coupling of \cref{thm:ent-max} with the allocation \eqref{eq:alloceuiv} to define the max-entropy rule.

\begin{defn}
The \emph{max-entropy rule} \(\left(\operatorname{ME}_{N}\right)_{N\geq 1}\) for \cref{prob:overlaps} is defined by
\[\operatorname{ME}_{N}:\Omega_{N}\longrightarrow\left[0,1\right]^{N},\quad \operatorname{ME}_{N}\left(\vec{w}\right)\coloneq\Phi_{N}\left(\mathbb{P}_{\vec{w}}^{\ast}\right)=\operatorname{Sh}\left(v_{\mathbb{P}_{\vec{w}}^{\ast}}\right).\]
Here \(\mathbb{P}_{\vec{w}}^{\ast}\) is given in \cref{thm:ent-max}, and \(\Phi_{N}\) and \(v_{\mathbb{P}}\) are defined in \S\ref{sct:alloc}, and \(\operatorname{Sh}\) stands for the Shapley value.
\end{defn}

Being compatible with the full-information benchmark is not assumed as an axiom in \cref{mainthm:shares}. Still, the resulting max-entropy rule is by definition compatible with the full-information benchmark.

\smallskip

On the part of each agent, the max-entropy rule is given by Diskin's rule. This yields explicit formulas, and shows that it extends Diskin's rule on the total association domain.

\begin{prop}\label{prop:ent-maxlloc}
Let \(\vec{w}\in\Omega_{N}\) with \(\sum_{i=1}^{N}w_{i}>1\), and put \(c\coloneq c_{\vec{w}}\). Then for every \(i\in\left[N\right]\),
\begin{equation}\label{eq:me-part}
\operatorname{ME}_{N,i}\left(\vec{w}\right)=w_{i}\operatorname{Disk}_{N,i}\left(1,c\vec{w}_{-i}\right).
\end{equation}
Explicitly, the max-entropy rule is given by the formula
\begin{equation}\label{eq:alloc-integral}
\operatorname{ME}_{N,i}\left(\vec{w}\right)=w_{i}\int_{0}^{1}\prod\nolimits_{j\in\left[N\right]\setminus\{i\}}\left(1-cw_{j}+cw_{j}t\right)dt=w_{i}\sum\nolimits_{J\subseteq\left[N\right]\setminus\left\{i\right\}}\mfrac{\left(-c\right)^{\left|J\right|}}{\left|J\right|+1}\prod\nolimits_{j\in J}w_{j}.
\end{equation}
In particular, the max-entropy rule extends Diskin's rule on the total association domain:
\begin{equation}\label{eq:me-extends-diskin}
\operatorname{ME}_{N}\left(\vec{w}\right)=\operatorname{Disk}_{N}\left(\vec{w}\right),\quad\vec{w}\in\Omega_{N}^{\mathrm{tot}}.
\end{equation}
\end{prop}

\begin{proof}
The case \(w_{i}=0\) is trivial, so assume \(w_{i}>0\). For \(T\subseteq\left[N\right]\) with \(i\in T\), the formula \eqref{eq:maxent-coupling} in \cref{thm:ent-max}(3) gives
\[\mathbb{P}_{\vec{w}}^{\ast}\left(T\right)=\mfrac{cw_{i}}{c}\Big(\prod\nolimits_{j\in T\setminus\left\{i\right\}}cw_{j}\Big)\Big(\prod\nolimits_{j\in\left[N\right]\setminus T}\left(1-cw_{j}\right)\Big)=w_{i}p_{T}^{\left(1,c\vec{w}_{-i}\right)}.\]
Therefore,
\[\operatorname{ME}_{N,i}\left(\vec{w}\right)=\sum\nolimits_{\left\{T\subseteq\left[N\right]:i\in T\right\}}\mfrac{\mathbb{P}_{\vec{w}}^{\ast}\left(T\right)}{\left|T\right|}=w_{i}\sum\nolimits_{\left\{T\subseteq\left[N\right]:i\in T\right\}}\mfrac{p_{T}^{\left(1,c\vec{w}_{-i}\right)}}{\left|T\right|}
=w_{i}\operatorname{Disk}_{N,i}\left(1,c\vec{w}_{-i}\right),\]
which proves \eqref{eq:me-part}. Applying \eqref{eq:diskin-integral} to \(\left(1,c\vec{w}_{-i}\right)\in\Omega_{N}^{\mathrm{tot}}\) gives \eqref{eq:alloc-integral}. On the total association domain, \(\mathbb{P}_{\vec{w}}^{\ast}=\mathbb{D}_{\vec{w}}\) by \cref{thm:ent-max}, so \eqref{eq:diskin-expectation} gives \eqref{eq:me-extends-diskin}.
\end{proof}

\subsection{Characterization by resource consistency}\label{sct:shares}

We can now reformulate \cref{mainthm:shares,mainthm:ent-max}.

\begin{thm}\label{thm:shares}
The max-entropy rule \(\left(\operatorname{ME}_{N}\right)_{N\geq 1}\) is the unique efficient, symmetric, null rule that satisfies resource consistency. For every  \(\vec{w}\in\Omega_{N}\) with \(\sum_{i=1}^{N}w_{i}>1\), the resource consistency factors all coincide, and their common value is the overlap factor:
\[c_{j}=c_{\vec{w}}\text{ whenever }0<w_{j}<1.\]
The rule extends concede-and-divide for two agents and Diskin's rule on the total association domain.
\end{thm}

The coincidence of the resource consistency factors is a major step in the proof of \cref{thm:shares}, and it is mainly due to the following technical lemma.

\begin{lem}\label{lem:factors}
Let \(\left(R_{N}\right)_{N\geq1}\) be an efficient, symmetric, null, resource consistent rule. Fix \(\vec{w}\in\Omega_{N}\), and suppose that \(R_{N}\left(\vec{u}\right)=\operatorname{ME}_{N}\left(\vec{u}\right)\) for every \(\vec{u}\in\Omega_{N}\) that has fewer positive coordinates than \(\vec{w}\). Then the resource consistency factor \(c_{j}\) for \(0<w_{j}<1\) is uniquely determined, and is given by
\[c_{j}=\begin{cases}
0&\sum_{i=1}^{N}w_{i}=1\\
c_{\vec{w}}&\sum_{i=1}^{N}w_{i}>1
\end{cases}.\]
\end{lem}

\begin{proof}
Put \(J\coloneq\left\{i\in\left[N\right]:w_{i}>0\right\}\).
If \(\left|J\right|=1\), the only positive coordinate equals \(1\), and there are no resource consistency factors to determine. Assume then that \(\left|J\right|\geq2\). When \(\sum_{i=1}^{N}w_{i}=1\), admissibility of the complementary part gives
\[1\leq c_{j}^{\ast}\sum\nolimits_{i\neq j}w_{i}=c_{j}^{\ast}\left(1-w_{j}\right)=1-c_{j}w_{j},\text{ hence }c_{j}=0.\]
Resource consistency then gives \(R_{N}\left(\vec{w}\right)=w_{j}R_{N}\big(1,\vec{0}\big)+\left(1-w_{j}\right)R_{N}\left(0,\vec{w}_{-j}/\left(1-w_{j}\right)\right)\). By nullity and efficiency, \(R_{N}\big(1,\vec{0}\big)=\big(1,\vec{0}\big)\). The second vector has fewer positive coordinates and its coordinates sum to \(1\), so by the assumption of the lemma and \cref{thm:ent-max}, \(R_{N}\left(\vec{w}\right)=\vec{w}\). 

\smallskip

Assume that \(\sum_{i=1}^{N}w_{i}>1\). On the total association domain we have \(c_{j}=c_{\vec{w}}=1\), as shown after \cref{dfn:resource-consistency}. Then consider the case \(\max_{i\in\left[N\right]}w_{i}<1\). For every \(j\in J\) we have \(c_{j}^{\ast}>0\) by admissibility, hence \(\left(0,c_{j}^{\ast}\vec{w}_{-j}\right)\) has \(\left|J\right|-1\) positive coordinates, so the assumption in the lemma applies to this vector.

\smallskip

Let us next show that all factors coincide. If \(\left|J\right|=2\), let \(j,k\) be its two elements. On the complementary part of \(j\), the only positive association is \(c_{j}^{\ast}w_{k}\), so \(c_{j}^{\ast}w_{k}=1\) by admissibility, and
\[c_{j}=\mfrac{w_{j}+w_{k}-1}{w_{j}w_{k}}.\]
This is symmetric in \(j\) and \(k\), and it is the nonzero solution of \(c=1-\left(1-cw_{j}\right)\left(1-cw_{k}\right)\). Thus \(c_{j}=c_{k}=c_{\vec{w}}\). Assume then that \(\left|J\right|\geq3\). For every \(k\in\left[N\right]\), put
\[F_{k}\left(c\right)\coloneq\int_{0}^{1}\prod\nolimits_{i\in\left[N\right]\setminus\left\{k\right\}}\left(1-cw_{i}+cw_{i}t\right)dt,\quad c\geq0.\]
Fix \(k\in J\). By the assumption \(R_{N,k}\left(0,c_{k}^{\ast}\vec{w}_{-k}\right)=\operatorname{ME}_{N,k}\left(0,c_{k}^{\ast}\vec{w}_{-k}\right)=0\), and by \cref{prop:diskin-char} we have \(R_{N,k}\left(1,c_{k}\vec{w}_{-k}\right)=\operatorname{Disk}_{N,k}\left(1,c_{k}\vec{w}_{-k}\right)\). Then by resource consistency and the formula \eqref{eq:diskin-integral},
\begin{equation}\label{eq:0th}
R_{N,k}\left(\vec{w}\right)=w_{k}R_{N,k}\left(1,c_{k}\vec{w}_{-k}\right)=w_{k}\operatorname{Disk}_{N,k}\left(1,c_{k}\vec{w}_{-k}\right)=w_{k}F_{k}\left(c_{k}\right).
\end{equation}

\smallskip

Fix some \(j\in J\setminus\left\{k\right\}\). Put
\[b_{j}\coloneq\begin{cases}
0
&\sum_{i\neq j}c_{j}^{\ast}w_{i}=1\\
c_{j}^{\ast}c_{c_{j}^{\ast}\vec{w}_{-j}}
&\sum_{i\neq j}c_{j}^{\ast}w_{i}>1
\end{cases}.\]
In Appendix~\ref{app:fact-comp} we calculate the identities
\begin{equation}\label{eq:6th}
\operatorname{sgn}\left(c_{j}-c_{k}\right)
=\operatorname{sgn}\left(F_{k}\left(c_{k}\right)-F_{k}\left(c_{j}\right)\right)=\operatorname{sgn}\left(c_{j}-b_{j}\right),\quad j,k\in J,\ j\neq k.
\end{equation}
We can now deduce the coincidence of all factors. Fix \(i_{\min},i_{\max}\in J\) such that \(c_{i_{\min}}=\min_{i\in J}c_{i}\) and \(c_{i_{\max}}=\max_{i\in J}c_{i}\). As \(\left|J\right|\geq3\), we may fix some \(j\in J\setminus\left\{i_{\min},i_{\max}\right\}\). Using \eqref{eq:6th} twice, we get
\[0\leq\operatorname{sgn}\left(c_{j}-c_{i_{\min}}\right)=\operatorname{sgn}\left(c_{j}-b_{j}\right)=\operatorname{sgn}\left(c_{j}-c_{i_{\max}}\right)\leq0.\]
Thus \(c_{i_{\min}}=c_{i_{\max}}\), and all factors coincide.

\smallskip

It remains to determine their common value \(c\). For \(k\notin J\), nullity gives \(R_{N,k}\left(\vec{w}\right)=0=w_{k}F_{k}\left(c\right)\). Hence \(R_{N,k}\left(\vec{w}\right)=w_{k}F_{k}\left(c\right)\) for every \(k\in\left[N\right]\). If \(c=0\) then \(F_{k}\left(c\right)=1\) so \(R_{N,k}\left(\vec{w}\right)=w_{k}F_{k}\left(c\right)=w_{k}\), contrary to efficiency and \(\sum_{i=1}^{N}w_{i}>1\). Thus \(c>0\). Using efficiency and the formula \eqref{eq:0th}, we get
\begin{align*}
c=c\sum\nolimits_{k=1}^{N}R_{N,k}\left(\vec{w}\right)
&=\int_{0}^{1}\sum\nolimits_{k=1}^{N}cw_{k}\prod\nolimits_{i\neq k}\left(1-cw_{i}+cw_{i}t\right)dt\\
&=\int_{0}^{1}\mfrac{d}{dt}\prod\nolimits_{i=1}^{N}\left(1-cw_{i}+cw_{i}t\right)dt
=1-\prod\nolimits_{i=1}^{N}\left(1-cw_{i}\right).
\end{align*}
This is the defining equation of the overlap factor in \cref{dfn:overfac}. The factors \(1-cw_{i}\) are nonnegative by admissibility, so \(0<c\leq1\), and uniqueness gives \(c=c_{\vec{w}}\).
\end{proof}

\begin{proof}[Proof of \cref{thm:shares}]
We first show that the max-entropy rule has the stated properties. It coincides with Diskin's rule on the total association domain by \eqref{eq:me-extends-diskin}. For two agents, admissibility determines a unique coupling, and equal division gives concede-and-divide, as shown in the introduction. It is efficient and null by \eqref{eq:alloceuiv}, and anonymous, hence symmetric, since the max-entropy construction is invariant under permutations of the agents.

\smallskip

We now look at resource consistency. When \(\sum_{i=1}^{N}w_{i}=1\), by \cref{thm:ent-max} the region associated with the \(i^{\mathrm{th}}\) agent alone has size \(w_{i}\), so \eqref{eq:alloceuiv} gives \(\operatorname{ME}_{N}\left(\vec{w}\right)=\vec{w}\). For a cut along \(j\) with \(0<w_{j}<1\), take \(c_{j}=0\). The reconstructed association vectors are \(\big(1,\vec{0}\big)\) and \(\left(0,\vec{w}_{-j}/\left(1-w_{j}\right)\right)\). Both vectors still have their coordinates sum to \(1\), so the rule returns these vectors on the two parts. Weighting them by \(w_{j}\) and \(1-w_{j}\) gives \(\vec{w}\), and thus resource consistency follows. When  \(\sum_{i=1}^{N}w_{i}>1\), put \(c\coloneq c_{\vec{w}}\) and take \(c_{j}=c\) for every \(j\in\left[N\right]\) with \(0<w_{j}<1\). Since the max-entropy coupling is consistent with cuts (\cref{prop:split}), applying \eqref{eq:benchmark-cut} to \(\mathbb{P}=\mathbb{P}_{\vec{w}}^{\ast}\) gives
\[\operatorname{ME}_{N}\left(\vec{w}\right)=w_{j}\operatorname{ME}_{N}\left(1,c\vec{w}_{-j}\right)+\left(1-w_{j}\right)\operatorname{ME}_{N}\left(0,c_{j}^{\ast}\vec{w}_{-j}\right),\quad c_{j}^{\ast}=\mfrac{1-cw_{j}}{1-w_{j}}.\]
Both vectors are in the domain, since \(0\leq cw_{k}\leq1\) for every \(k\neq j\), and \(c_{j}^{\ast}\vec{w}_{-j}\in\Omega_{N-1}\) by \cref{prop:split}.

\smallskip

Conversely, let \(\left(R_{N}\right)_{N\geq1}\) be an efficient, symmetric, null, resource consistent rule, and fix \(N\geq1\). We prove both assertions by induction on the number of positive coordinates of \(\vec{w}\in\Omega_{N}\). If only one coordinate is positive then it equals \(1\), so efficiency and nullity give \(R_{N}\left(\vec{w}\right)=\vec{w}=\operatorname{ME}_{N}\left(\vec{w}\right)\), and there are no resource consistency factors to determine. Assume that \(\vec{w}\) has at least two positive coordinates, and assume by induction that \(R_{N}\left(\vec{u}\right)=\operatorname{ME}_{N}\left(\vec{u}\right)\) for every \(\vec{u}\in\Omega_{N}\) with fewer positive coordinates. By \cref{lem:factors}, all resource consistency factors of \(\vec{w}\) are uniquely determined as stated there.

\smallskip

When \(\sum_{i=1}^{N}w_{i}=1\), we have \(R_{N}\left(\vec{w}\right)=\vec{w}=\operatorname{ME}_{N}\left(\vec{w}\right)\) as shown at the beginning of the proof of \cref{lem:factors}. Assume then that \(\sum_{i=1}^{N}w_{i}>1\), so every factor equals \(c_{\vec{w}}\). On the total association domain, \(R_{N}=\operatorname{Disk}_{N}=\operatorname{ME}_{N}\) by \cref{prop:diskin-char} and \eqref{eq:me-extends-diskin}. It remains to consider the case \(\max_{i\in\left[N\right]}w_{i}<1\). For every \(j\) with \(w_{j}>0\), nullity and resource consistency, together with \cref{prop:ent-maxlloc}, give
\[R_{N,j}\left(\vec{w}\right)=w_{j}R_{N,j}\left(1,c_{\vec{w}}\vec{w}_{-j}\right)=w_{j}\operatorname{Disk}_{N,j}\left(1,c_{\vec{w}}\vec{w}_{-j}\right)=\operatorname{ME}_{N,j}\left(\vec{w}\right).\]
For \(w_{j}=0\), both shares vanish by nullity. Hence \(R_{N}\left(\vec{w}\right)=\operatorname{ME}_{N}\left(\vec{w}\right)\), and the induction is complete.
\end{proof}

\section{Contested ownership and claim rules}\label{sct:claims}

We now return to contested ownership (\cref{prob:Ncontgarm}), where \emph{claims} play the role of \emph{associations}.

\subsection{The method of random claims and its flaw}\label{sct:randomclaims}

\citet{ONeill1982} studied several methods for the simple claims problem, one of which is the \emph{method of random claims} \citep[\S5.1]{ONeill1982}. In this method, every infinitesimal unit of the object is claimed by a random set of claimants and is divided equally among them, where the law of this set is determined by requiring that the \(i^{\mathrm{th}}\) claimant claims a random unit with probability \(w_{i}\), independently of the other claimants. In our notation, \(S\) is the set of claimants of a random unit, and the law of the random claims is the independent coupling \(\mathbb{D}_{\vec{w}}\) of \eqref{eq:indep}.

\smallskip

For \(\vec{w}\in\left[0,1\right]^{N}\setminus\big\{\vec{0}\big\}\) put
\[p_{\vec{w}}\coloneq\mathbb{D}_{\vec{w}}\left(S\neq\emptyset\right)=1-\prod\nolimits_{i=1}^{N}\left(1-w_{i}\right).\]
Unless some claim is total, \(p_{\vec{w}}<1\), so the first round leaves part of the object unclaimed. Then the procedure is repeated on the remainder until the entire object is covered. In each round, a proportion \(\mathbb{D}_{\vec{w}}\left(T\right)\) of the available part is claimed exactly by the members of \(T\), and a proportion \(1-p_{\vec{w}}\) of it remains unclaimed. Hence the total proportion eventually claimed exactly by the members of \(T\in\mathcal{S}_{N}\) is
\[\sum\nolimits_{m=0}^{\infty}\left(1-p_{\vec{w}}\right)^{m}\mathbb{D}_{\vec{w}}\left(T\right)=\mfrac{\mathbb{D}_{\vec{w}}\left(T\right)}{p_{\vec{w}}}=\mathbb{D}_{\vec{w}}\left(T\mid S\neq\emptyset\right).\]
Thus the overlap model produced by the method is \(\mathbb{D}_{\vec{w}}\left(\cdot\mid S\neq\emptyset\right)\), with marginal proportions \(w_{i}/p_{\vec{w}}\). When no claim is total, these exceed the prescribed proportions and may violate claims boundedness, which O'Neill identified as a `serious flaw' \citep[\S5.1]{ONeill1982}; see also \citet[p.~44]{Thomson2019}.

\smallskip

Let us record the awards of the method as a rule. The method makes sense for every \(\vec{w}\in\left[0,1\right]^{N}\setminus\{\vec{0}\}\), since \(p_{\vec{w}}>0\) as soon as some claim is positive, and its awards are
\begin{equation}\label{eq:rc-rule}
\operatorname{RC}_{N,i}\left(\vec{w}\right)\coloneq\Phi_{N,i}\big(\mathbb{D}_{\vec{w}}\left(\cdot\mid S\neq\emptyset\right)\big)=\mfrac{1}{p_{\vec{w}}}\sum\nolimits_{\left\{T\in\mathcal{S}_{N}:i\in T\right\}}\mfrac{p_{T}^{\vec{w}}}{\left|T\right|},\quad i\in\left[N\right],
\end{equation}
where \(p_{T}^{\vec{w}}\) is defined in \eqref{eq:diskin-regions}. On the total association domain \(p_{\vec{w}}=1\) and \eqref{eq:rc-rule} is Diskin's rule. The method of random claims is an instance of the Shapley value of a probability game, introduced by
\citet{Hou2018}, in which the value of a coalition is the probability that at least one of its members succeeds when the members succeed independently. \citet{Dehez2023} gave an axiomatic foundation to its Shapley value. With the claims as the probabilities of success, the awards \eqref{eq:rc-rule} are proportional to the Shapley value of this game \citep[eq.~(9)]{Dehez2024}.

\smallskip

By \cref{thm:ent-max}(3), when \(\sum_{i=1}^{N}w_{i}>1\) we have \(\mathbb{P}_{\vec{w}}^{\ast}=\mathbb{D}_{c_{\vec{w}}\vec{w}}\left(\cdot\mid S\neq\emptyset\right)\), and by the polynomial equation defining the overlap factor in \cref{dfn:overfac},
\[p_{c_{\vec{w}}\vec{w}}=\mathbb{D}_{c_{\vec{w}}\vec{w}}\left(S\neq\emptyset\right)=1-\prod\nolimits_{i=1}^{N}\left(1-c_{\vec{w}}w_{i}\right)=c_{\vec{w}}.\]
Comparing with \eqref{eq:rc-rule} we obtain that for every \(\vec{w}\in\Omega_{N}\) with \(\sum\nolimits_{i=1}^{N}w_{i}>1\),
\begin{equation}\label{eq:me-rc}
\operatorname{ME}_{N}\left(\vec{w}\right)=\operatorname{RC}_{N}\left(c_{\vec{w}}\vec{w}\right).
\end{equation}
This recalibration is what makes the final distribution respect the original claimed proportions, so the flaw disappears. Accordingly, the awards of the max-entropy rule are proportional to the Shapley value of the probability game with the probabilities of success \(c_{\vec{w}}w_{i}\), and expanding this Shapley value as in \citet[eq.~(8)]{Dehez2024} gives the integration-free formula in \eqref{eq:alloc-integral}. For \(\vec{w}\in\Omega_{N}^{\mathrm{tot}}\) with \(\sum_{i=1}^{N}w_{i}>1\), we have \(c_{\vec{w}}=1\), and \eqref{eq:me-rc} reduces to \eqref{eq:me-extends-diskin}.

\smallskip

The flaw in the method of random claims can be repaired by the \emph{random stakes rule} of \citet[\S2.1.10]{Thomson2019}, who attributed its definition to \citet{ONeill1982}, that revises the claims iteratively. At every stage, every claimant claims each unit of the part which is still available with probability which is the ratio of his remaining claim and the size of this part, or with probability \(1\) if his remaining claim is larger. The claims at every stage are independent of the other claimants. After each unit is divided equally among its claimants, every claim is reduced by the award the claimant received. The procedure is repeated on the unclaimed part. The resulting rule is well defined \citep[Prop.~2.2]{Thomson2019}, and it is claims bounded, efficient, and anonymous. It also can be seen that it extends concede-and-divide for two claimants. On the total association domain, the first round exhausts the object and we get
\[\operatorname{RS}_{N}\left(\vec{w}\right)=\operatorname{Disk}_{N}\left(\vec{w}\right)=\operatorname{ME}_{N}\left(\vec{w}\right).\]
As we shall see in Appendix~\ref{app:suppfinf}, the random stakes rule is not full-information benchmark compatible.

\subsection{Comparison with other claims rules}\label{sct:comparison}

The simplest conceivable rule for \cref{prob:Ncontgarm}, which does not extend concede-and-divide for two claimants, is the proportional rule:
\[\operatorname{PROP}_{N,i}\left(\vec{w}\right)\coloneq\mfrac{w_{i}}{\sum\nolimits_{j=1}^{N}w_{j}},\quad i\in\left[N\right].\]

\smallskip

A rule \(\left(R_{N}\right)_{N\geq1}\) is \emph{bilaterally consistent} if whenever all but two claimants leave with their awards, it assigns the same awards to the remaining pair in the reduced problem \citep[\S10.2]{Thomson2019}.

\smallskip

To speak of bilateral consistency, we extend the max-entropy rule to a resource of size \(E>0\) in a standard way; see \citep[\S2.3]{Moulin2003}. Thus for a claim vector \(\vec{v}\in\left[0,\infty\right)^{N}\) with \(\sum_{i=1}^{N}v_{i}\geq E\), define
\[\operatorname{ME}_{N}\left(\vec{v},E\right)\coloneq E\cdot\operatorname{ME}_{N}\big(\min\big\{v_{1}/E,1\big\},\dotsc,\min\big\{v_{N}/E,1\big\}\big).\]
This also defines the rule on reduced problems in which a remaining claim exceeds the remaining estate. For \(E=0\) we put \(\operatorname{ME}_{N}\left(\vec{v},0\right)\coloneq\vec{0}\).

\smallskip

We consider two properties on the domain \(\Omega_{N}\). \emph{Strict cross-claim monotonicity} means that increasing one claim reduces the award of every other claimant with a positive claim. \emph{Strict population monotonicity} means that adding a claimant with a positive claim reduces the award of every existing claimant with a positive claim. The formal definitions and the proofs for the max-entropy rule are in Appendix~\ref{app:properties}.

\smallskip

There are several established extensions of concede-and-divide. The first is the \emph{Talmud rule}, which is the unique bilaterally consistent extension of concede-and-divide \citep{AumannMaschler1985}; see also \citet[Prop.~10.6]{Thomson2019}. Another such rule is the \emph{minimal overlap rule}, proposed by O'Neill as the \emph{extended Ibn Ezra method}. It arranges the claims in a way that lexicographically maximizes the measures of the regions claimed exactly once, exactly twice, and so on, and finally divides each region equally among its claimants \citep[\S2.1.9]{Thomson2019}. The more recent \emph{average-of-awards rule} places the uncertainty on the awards rather than on the claims: it selects the uniform distribution on the set of awards vectors of the problem and awards its expectation \citep{Miras2022}. It extends concede-and-divide and it is not bilaterally consistent \citep[Prop.~4.3]{Miras2022}. The examples in \cref{exm:diskin,exm:minimal-overlap} show that the average-of-awards rule fails the two strict properties satisfied by the max-entropy rule. For a thorough comparison of such rules, see \citet{Miras2026}.

\smallskip

A further extension of concede-and-divide is the \emph{random arrival rule} \citep[\S2.1.8]{Thomson2019}, equivalent to \emph{method of recursive completion} \citep{ONeill1982}, where claimants arrive in a uniformly random order and each is paid in full as long as the object lasts. It extends concede-and-divide \citep[p.~38]{Thomson2019}. \citet[Exercise~2.10]{Moulin2003} presents this procedure under the name ``run to the bank.'' It is the Shapley value of the usual coalitional game associated with a claims vector \(\vec{w}\in\Omega_{N}\),
\[v_{\vec{w}}^{\mathrm{cl}}\left(A\right)\coloneq\max\big\{1-\sum\nolimits_{i\notin A}w_{i},0\big\},\quad A\subseteq\left[N\right];\]
see \citet[\S14.1.2, Thm.~14.2(a)]{Thomson2019}. By contrast, the max-entropy rule is the Shapley value of the game induced by the max-entropy coupling.

\begin{exm}[Diskin's example]\label{exm:diskin}
For two claimants, the concede-and-divide rule gives
\[\operatorname{CD}\left(1,1/2\right)=\left(3/4,1/4\right).\]
When a third claimant with claim \(1/2\) joins, the Talmud rule gives
\[\operatorname{Talmud}_{3}\left(1,1/2,1/2\right)=\left(1/2,1/4,1/4\right),\]
so the second claimant's award remains \(1/4\), and strict population monotonicity fails. When the second claim is increased to \(1\), the Talmud rule gives
\[\operatorname{Talmud}_{3}\left(1,1,1/2\right)=\left(3/8,3/8,1/4\right),\]
so strict cross-claim monotonicity also fails. By contrast, Diskin's division of \S\ref{sct:diskin-argument} gives
\[\operatorname{ME}_{3}\left(1,1/2,1/2\right)=\operatorname{Disk}_{3}\left(1,1/2,1/2\right)=\left(7/12,5/24,5/24\right)\]
and
\[\operatorname{ME}_{3}\left(1,1,1/2\right)=\operatorname{Disk}_{3}\left(1,1,1/2\right)=\left(5/12,5/12,1/6\right).\]
For \(\vec{w}=\left(1,1/2,1/2\right)\), the Talmud rule aligns with the overlap model of disjoint half-claims, and Diskin's rule aligns with the overlap model of one quarter. For \(\vec{w}=\left(1,1,1/2\right)\), there is a unique admissible coupling, and thus the Talmud rule does not follow any overlap model; indeed, this demonstrates the fact that the Talmud rule is not compatible with the full-information benchmark.
\end{exm}

\begin{exm}\label{exm:minimal-overlap}
For \(0\leq t\leq1\) let \(\vec{w}\left(t\right)\coloneq\left(1,1/2,t\right)\). We present in \cref{fig:secawards} the second claimant's award for this vector as a function of \(t\). The computations are given in Appendix~\ref{app:suppwt}. Since the first claim is total, the method of random claims and the random stakes rule coincide with the max-entropy rule.

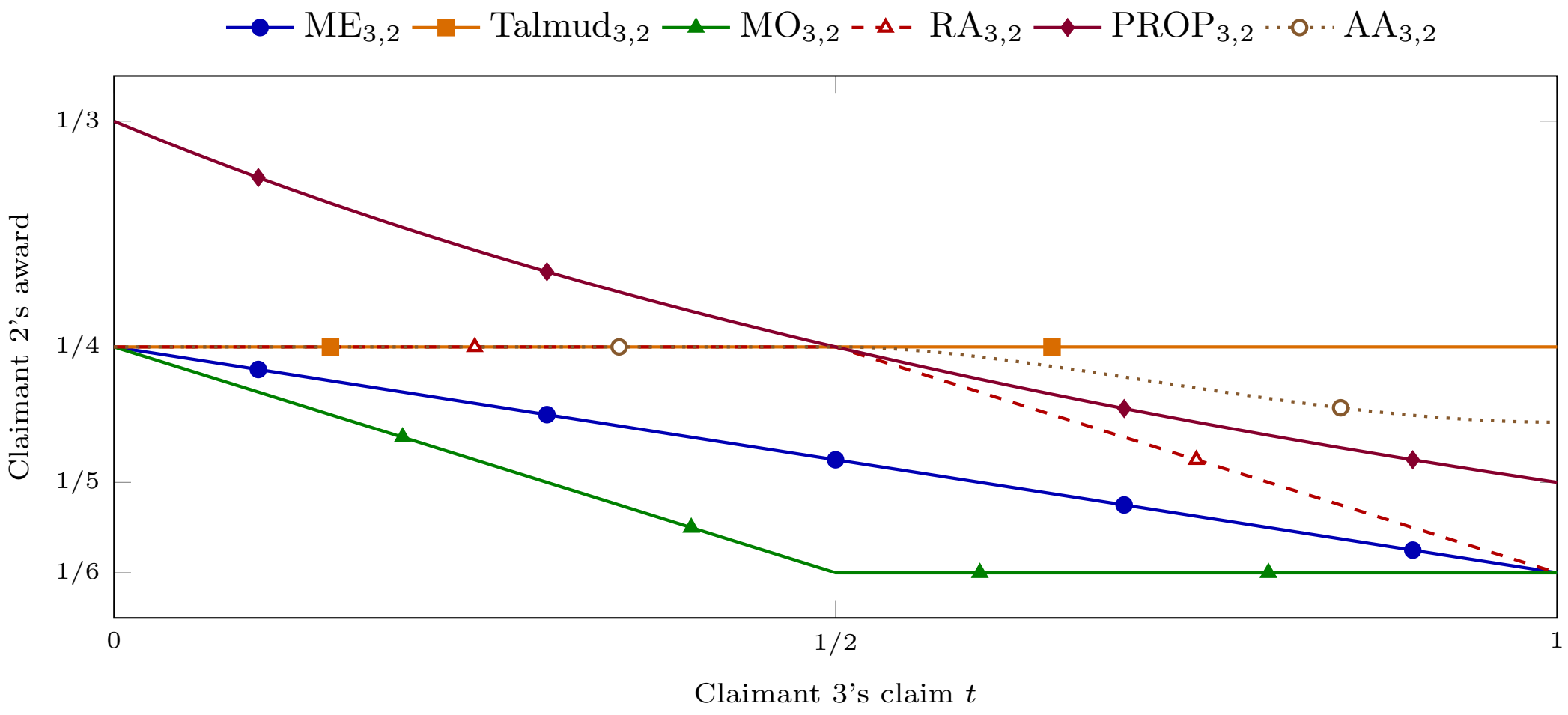
\begin{figure}[H]
\captionsetup{font=small}
\centering
\begin{tikzpicture}
\begin{axis}[
    width=0.9\textwidth,
    height=0.4\textwidth,
    xmin=0,
    xmax=1,
    ymin=0.15,
    ymax=0.35,
    xlabel={\scriptsize Claimant \(3\)'s claim \(t\)},
    ylabel={\scriptsize Claimant \(2\)'s award},
    xtick={0,0.5,1},
    xticklabels={\(0\),\(1/2\),\(1\)},
    ytick={0.1666666667,0.2,0.25,0.3333333333},
    yticklabels={\(1/6\),\(1/5\),\(1/4\),\(1/3\)},
    tick label style={font=\tiny},
    label style={font=\tiny},
    axis lines=box,
    domain=0:1,
    samples=101,
    every axis plot/.append style={thick,mark size=1.8pt},
    legend image post style={mark indices={2}},
    legend style={at={(0.5,1.03)},anchor=south,legend columns=6,draw=none,font=\small}
]
\addplot[blue!70!black,mark=*,mark indices={11,31,51,71,91}] {1/4-x/12};
\addlegendentry{\(\operatorname{ME}_{3,2}\)}
\addplot[orange!85!black,mark=square*,mark indices={16,66}] {1/4};
\addlegendentry{\(\operatorname{Talmud}_{3,2}\)}
\addplot[green!50!black,mark=triangle*,mark indices={21,41,61,81}] {max(1/4-x/6,1/6)};
\addlegendentry{\(\operatorname{MO}_{3,2}\)}
\addplot[red!70!black,dashed,mark=triangle*,mark options={solid,fill=white},mark indices={26,76}] {min(1/4,(2-x)/6)};
\addlegendentry{\(\operatorname{RA}_{3,2}\)}
\addplot[purple!70!black,mark=diamond*,mark indices={11,31,71,91}] {1/(3+2*x)};
\addlegendentry{\(\operatorname{PROP}_{3,2}\)}
\addplot[brown!70!black,dotted,mark=*,mark options={solid,fill=white},mark indices={36,86}] {x<=0.5 ? 1/4 : (4*x^3-12*x^2+12*x-2)/(3*(8*x-4*x^2-1))};
\addlegendentry{\(\operatorname{AA}_{3,2}\)}
\end{axis}
\end{tikzpicture}
\caption{Awards to the second claimant for \(\left(1,1/2,t\right)\).}
\label{fig:secawards}
\end{figure}
\end{exm}

All rules in \cref{fig:secawards} are efficient, claims bounded, and anonymous; see \citet[Ch.~3]{Thomson2019}, and for the average-of-awards rule see \citet{Miras2022}. The random stakes rule is efficient, claims bounded, and anonymous (\S\ref{sct:randomclaims}). The table compares only the properties on which the rules differ.

\begin{table}[H]
\captionsetup{font=footnotesize}
\centering
\footnotesize
\setlength{\tabcolsep}{4pt}
\begin{tabular}{lccccccc}
\toprule
Property \textbackslash{} Rule &
\makecell{Max-entropy} & Talmud & \makecell{Minimal\\overlap} & \makecell{Random\\arrival} & \makecell{Random\\stakes} & \makecell{Average\\of awards} & Proportional \\
\midrule
Extending concede-and-divide & Yes & Yes & Yes & Yes & Yes & Yes & No \\
Extending Diskin's rule & Yes & No & No & No & Yes & No & No \\
Full-info. benchmark compatibility & Yes & No & Yes & No & No & No & No \\
Resource consistency & Yes & No & No & No & No & No & No \\
Bilateral consistency & No & Yes & No & No & No & No & Yes \\
Strict cross-claim monotonicity & Yes & No & No & No & --- & No & Yes \\
Strict population monotonicity & Yes & No & Yes & No & --- & No & Yes \\
\bottomrule
\end{tabular}
\caption{Comparison of the max-entropy rule with other claims rules.}
\label{tab:comparison}
\end{table}

Every rule in the table other than the proportional rule extends concede-and-divide and differs from the Talmud rule (\cref{fig:secawards}), so by \citet[Prop.~10.6, p.~257]{Thomson2019} none of them is bilaterally consistent. The entries for the max-entropy rule follow from \cref{thm:shares,prop:properties}. Since all rules are efficient, symmetric and null, \cref{thm:shares} gives the resource consistency row. The failures of the strict properties are illustrated in \cref{exm:diskin,exm:minimal-overlap}. For the remaining entries, see \citet[\S\S2.1.8--2.1.9, \S10.2, \S12.1]{Thomson2019}, \citet[proof of Prop.~C.2]{Miras2023}, and \citet[proof of Prop.~4.3]{Miras2022}. The proportional rule entries follow directly. A dash (---) indicates a property not established here. The row on extending Diskin's rule follows from \eqref{eq:me-extends-diskin}, \S\ref{sct:randomclaims}, and the formulas in Appendix~\ref{app:suppwt} at \(t=1/2\). The entries on full-information benchmark compatibility are derived in Appendix~\ref{app:suppfinf}.

\appendix
\addcontentsline{toc}{section}{Appendices}
\addtocontents{toc}{\protect\setcounter{tocdepth}{0}}

\section{The overlap factor and its properties}\label{app:ovfac}

Here we justify the definition and basic properties of the overlap factor in \cref{dfn:overfac}.

\smallskip

Fix \(\vec{w}\in\Omega_{N}\) with \(\sum_{i=1}^{N} w_{i}>1\), and define
\[g_{\vec{w}}\left(c\right)\coloneq1-\prod\nolimits_{i=1}^{N}\left(1-cw_{i}\right)-c,\quad c\in\left[0,1\right].\]
Differentiation by the product rule gives
\[g_{\vec{w}}'\left(0\right)=\sum\nolimits_{i=1}^{N}w_{i}-1>0\] and
\[g_{\vec{w}}''\left(c\right)=-\sum\nolimits_{i=1}^{N}\sum\nolimits_{j\in\left[N\right]\setminus\{i\}}w_{i}w_{j}\prod\nolimits_{k\in\left[N\right]\setminus\{i,j\}}\left(1-cw_{k}\right)<0,\quad c\in\left[0,1\right).\]
Thus \(g_{\vec{w}}\) is strictly concave on \(\left[0,1\right]\). Since \(g_{\vec{w}}\left(0\right)=0\), it is positive immediately to the right of \(0\), while
\[g_{\vec{w}}\left(1\right)=-\prod\nolimits_{i=1}^{N}\left(1-w_{i}\right)\leq0.\]
Therefore it has a unique zero \(c_{\vec{w}}\in\left(0,1\right]\), and it is positive on \(\left(0,c_{\vec{w}}\right)\) and negative on \(\left(c_{\vec{w}},1\right]\). In particular, \(c_{\vec{w}}=1\) if and only if \(\max_{i\in\left[N\right]}w_{i}=1\).

\begin{lem}\label{lem:overfac}
The overlap factor \(c_{\vec{w}}\) presented in \cref{dfn:overfac} satisfies the following properties.
\begin{enumerate}
    \item For all \(\vec{w},\vec{w}^{\prime}\in\Omega_{N}\) with \(\sum_{i=1}^{N}w_{i}>1\), and for every \(j\) with \(w_{j}^{\prime}>w_{j}\) and \(w_{i}^{\prime}=w_{i}\), \(i\neq j\),
    \[c_{\vec{w}^{\prime}}\geq c_{\vec{w}}.\]
    \item For every \(\vec{w}\in\Omega_{N}\) with \(\sum_{i=1}^{N}w_{i}>1\) and every \(u\in\left(0,1\right]\),
    \[c_{\left(\vec{w},u\right)}\geq c_{\vec{w}}.\]
\end{enumerate}
\end{lem}

\begin{proof}
For (1) and (2), put \(c\coloneq c_{\vec{w}}>0\). Since \(g_{\vec{w}}\left(c\right)=0\), if \(w_{j}^{\prime}>w_{j}\) then
\[g_{\vec{w}^{\prime}}\left(c\right)=g_{\vec{w}^{\prime}}\left(c\right)-g_{\vec{w}}\left(c\right)=c\left(w_{j}^{\prime}-w_{j}\right)\prod\nolimits_{i\neq j}\left(1-cw_{i}\right)\geq0.\]
Since \(g_{\vec{w}^{\prime}}\) is nonnegative only on \(\left[0,c_{\vec{w}^{\prime}}\right]\), we get that \(c_{\vec{w}^{\prime}}\geq c\), proving (1). Similarly, if \(u\in\left(0,1\right]\) then
\[g_{\left(\vec{w},u\right)}\left(c\right)=g_{\left(\vec{w},u\right)}\left(c\right)-g_{\vec{w}}\left(c\right)=cu\prod\nolimits_{i=1}^{N}\left(1-cw_{i}\right)=cu\left(1-c\right)\geq0.\]
Since \(g_{\left(\vec{w},u\right)}\) is nonnegative only on \(\left[0,c_{\left(\vec{w},u\right)}\right]\), we get that \(c_{\left(\vec{w},u\right)}\geq c\), proving (2).
\end{proof}

\subsection{Comparison of two cuts}\label{app:fact-comp}

Here we prove the identities \eqref{eq:6th} regarding the overlap factor, which are used in the proof of \cref{lem:factors}. We continue using the notation and assumptions of that proof in the case \(\sum_{i=1}^{N}w_{i}>1\), \(\max_{i\in\left[N\right]}w_{i}<1\), and \(\left|J\right|\geq3\).

\smallskip

Let \(k\in J\). Fix some \(j\in J\setminus\left\{k\right\}\). For the part associated with \(j\), we use \cref{prop:diskin-char}. For its complement, we use the assumption in the lemma and \cref{prop:ent-maxlloc}. For the first computation, put
\[A_{j,k}^{c}\left(t\right)\coloneq\prod\nolimits_{i\in\left[N\right]\setminus\left\{j,k\right\}}\left(1-cw_{i}+cw_{i}t\right),\quad c\geq0.\]
Using \cref{prop:diskin-char} and the formula \eqref{eq:diskin-integral}, we may write
\begin{equation}\label{eq:1st}
\begin{aligned}
R_{N,k}\left(1,c_{j}\vec{w}_{-j}\right)
&=\operatorname{Disk}_{N,k}\left(1,c_{j}\vec{w}_{-j}\right)\\
&=c_{j}w_{k}\int_{0}^{1}\left(1-1+t\right)
\prod\nolimits_{i\in\left[N\right]\setminus\left\{j,k\right\}}
\left(1-c_{j}w_{i}+c_{j}w_{i}t\right)dt=c_{j}w_{k}\int_{0}^{1}tA_{j,k}^{c_{j}}\left(t\right)dt.
\end{aligned}
\end{equation}
By the assumption in the lemma, \cref{prop:ent-maxlloc}, and the formula \eqref{eq:alloc-integral}, we may write
\begin{equation}\label{eq:2nd}
\begin{aligned}
R_{N,k}\left(0,c_{j}^{\ast}\vec{w}_{-j}\right)
&=\operatorname{ME}_{N,k}\left(0,c_{j}^{\ast}\vec{w}_{-j}\right)\\
&=c_{j}^{\ast}w_{k}\int_{0}^{1}
\prod\nolimits_{i\in\left[N\right]\setminus\left\{j,k\right\}}
\left(1-b_{j}w_{i}+b_{j}w_{i}t\right)dt=c_{j}^{\ast}w_{k}\int_{0}^{1}A_{j,k}^{b_{j}}\left(t\right)dt.
\end{aligned}
\end{equation}
When \(\sum_{i\neq j}c_{j}^{\ast}w_{i}=1\), this follows from \(\operatorname{ME}_{N-1}\left(c_{j}^{\ast}\vec{w}_{-j}\right)=c_{j}^{\ast}\vec{w}_{-j}\).

\smallskip

Resource consistency for the cut along \(j\) means that \eqref{eq:0th} equals the sum of \eqref{eq:1st} and \eqref{eq:2nd} with appropriate weights. Then after dividing by \(w_{k}\) and using \(\left(1-w_{j}\right)c_{j}^{\ast}=1-c_{j}w_{j}\), it yields the formula
\begin{equation}\label{eq:3rd}
F_{k}\left(c_{k}\right)=c_{j}w_{j}\int_{0}^{1}tA_{j,k}^{c_{j}}\left(t\right)dt+\left(1-c_{j}w_{j}\right)\int_{0}^{1}A_{j,k}^{b_{j}}\left(t\right)dt.
\end{equation}

\smallskip

Let us now look at \(F_{k}\left(c_{j}\right)\). Separating the \(j^{\mathrm{th}}\) factor in its defining product, we may write
\begin{equation}\label{eq:4th}
F_{k}\left(c_{j}\right)=\int_{0}^{1}\left(1-c_{j}w_{j}+c_{j}w_{j}t\right)
A_{j,k}^{c_{j}}\left(t\right)dt=c_{j}w_{j}\int_{0}^{1}tA_{j,k}^{c_{j}}\left(t\right)dt
+\left(1-c_{j}w_{j}\right)\int_{0}^{1}A_{j,k}^{c_{j}}\left(t\right)dt.
\end{equation}
Subtracting \eqref{eq:4th} from \eqref{eq:3rd} yields
\begin{equation}\label{eq:5th}
F_{k}\left(c_{k}\right)-F_{k}\left(c_{j}\right)=\left(1-c_{j}w_{j}\right)\int_{0}^{1}\big(A_{j,k}^{b_{j}}\left(t\right)-A_{j,k}^{c_{j}}\left(t\right)\big)dt.
\end{equation}

Let \(I_{j,k}\) be the interval whose endpoints are \(c_{j}\) and \(c_{k}\). We claim that \(F_{k}\) is strictly decreasing on \(I_{j,k}\). Indeed, admissibility gives \(0\leq c_{k}w_{i}\leq1\) for \(i\neq k\) and \(0\leq c_{j}w_{i}\leq1\) for \(i\neq j\). Also for \(i=j\) we have
\[0\leq c_{j}w_{j}=1-\left(1-w_{j}\right)c_{j}^{\ast}\leq1.\]
Thus \(0\leq cw_{i}\leq 1\) for every \(c\in I_{j,k}\) and \(i\in\left[N\right]\setminus\{k\}\). Hence, for every \(0<t<1\) and \(i\in\left[N\right]\setminus\{k\}\),
\[c\mapsto1-cw_{i}+cw_{i}t=1-\left(1-t\right)cw_{i}\]
is positive and nonincreasing on \(I_{j,k}\), and is strictly decreasing if \(i\in J\). Since \(J\setminus\left\{k\right\}\neq\emptyset\), this implies that \(F_{k}\) is strictly decreasing on \(I_{j,k}\), and hence
\[\operatorname{sgn}\left(F_{k}\left(c_{k}\right)-F_{k}\left(c_{j}\right)\right)=\operatorname{sgn}\left(c_{j}-c_{k}\right).\]
A similar reasoning applies to show that for every \(0<t<1\), the function \(c\mapsto A_{j,k}^{c}\left(t\right)\) is strictly decreasing on the interval whose endpoints are \(b_{j}\) and \(c_{j}\). Consequently, using \eqref{eq:5th} we obtain \eqref{eq:6th}.

\section{The max-entropy coupling}\label{app:entopt}

Here we prove the implication \((2)\Rightarrow(3)\) in \cref{thm:ent-max} for \(\max_{i\in\left[N\right]}w_{i}<1\). We use the following fact about conditioned product distributions. Recall the notation \eqref{eq:indep}.

\begin{lem}\label{lem:cond-prod}
Let \(\mathbb{P}\in\operatorname{Prob}\left(\mathcal{S}_{N}\right)\) be fully supported, and suppose that for every \(i\in\left[N\right]\), the indicators \(X_{j}\), \(j\in\left[N\right]\setminus\{i\}\), are mutually independent under \(\mathbb{P}\left(\cdot\mid X_{i}=1\right)\). Then there exists \(\vec{r}\in\left(0,1\right)^{N}\) such that
\[\mathbb{P}=\mathbb{D}_{\vec{r}}\left(\cdot\mid S\neq\emptyset\right).\]
\end{lem}

\begin{proof}
Fix \(j\in\left[N\right]\) and \(\emptyset\neq T\subseteq\left[N\right]\setminus\left\{j\right\}\), and choose some \(i\in T\). By the assumption, \(X_{j}\) and \(S\setminus\left\{j\right\}\) are independent under \(\mathbb{P}\left(\cdot\mid X_{i}=1\right)\). Since both \(T\) and \(\left[N\right]\setminus\left\{j\right\}\) contain \(i\), we have
\[\mfrac{\mathbb{P}\left(T\cup\left\{j\right\}\right)}{\mathbb{P}\left(T\right)}
=\mfrac{\mathbb{P}\left(X_{j}=1\mid X_{i}=1\right)}{\mathbb{P}\left(X_{j}=0\mid X_{i}=1\right)}
=\mfrac{\mathbb{P}\left(\left[N\right]\right)}{\mathbb{P}\left(\left[N\right]\setminus\left\{j\right\}\right)}\coloneq a_{j}.\]
By full support, \(a_{j}>0\). For \(T\in\mathcal{S}_{N}\) with \(T\neq\left[N\right]\), enumerate \(\left[N\right]\setminus T=\left\{i_{1},\dotsc,i_{m}\right\}\). Then iteratively,
\[\mathbb{P}\left(\left[N\right]\right)
=a_{i_{1}}\mathbb{P}\left(\left[N\right]\setminus\left\{i_{1}\right\}\right)
=\dots=a_{i_{1}}\dotsm a_{i_{m}}\mathbb{P}\left(T\right)
=\mathbb{P}\left(T\right)\prod\nolimits_{i\in\left[N\right]\setminus T}a_{i}.\]
For \(T=\left[N\right]\) this means \(\prod_{i\in\left[N\right]\setminus T}a_{i}=1\). Put \(r_{i}\coloneq a_{i}/\left(1+a_{i}\right)\) for \(i\in\left[N\right]\). Then for every \(T\in\mathcal{S}_{N}\),
\[\mfrac{\mathbb{P}\left(T\right)}{\mathbb{P}\left(\left[N\right]\right)}
=\prod\nolimits_{i\in\left[N\right]\setminus T}\mfrac{1}{a_{i}}
=\prod\nolimits_{i\in\left[N\right]\setminus T}\mfrac{1-r_{i}}{r_{i}}
=\mfrac{\mathbb{D}_{\vec{r}}\left(T\right)}{\mathbb{D}_{\vec{r}}\left(\left[N\right]\right)}.\]
Thus \(\mathbb{D}_{\vec{r}}\left(T\right)\) is proportional to \(\mathbb{P}\left(T\right)\) on \(\mathcal{S}_{N}\). Conditioning on \(S\neq\emptyset\) gives \(\mathbb{P}\).
\end{proof}

\begin{proof}[Proof of \((2)\Rightarrow(3)\) in \cref{thm:ent-max}]
Assume that \(\sum_{i=1}^{N}w_{i}>1\) and \(\max_{i\in\left[N\right]}w_{i}<1\). As in the proof of \cref{thm:ent-max}, we may assume that \(w_{i}>0\) for every \(i\). Let \(\mathbb{P}\in\mathcal{P}_{N}\left(\vec{w}\right)\) satisfy (2). Since \(\mathbb{D}_{\vec{w}}\left(T\right)>0\) for every \(T\in\mathcal{S}_{N}\), \(\mathbb{P}\) has full support on \(\mathcal{S}_{N}\). By \cref{lem:cond-prod}, \(\mathbb{P}=\mathbb{D}_{\vec{r}}\left(\cdot\mid S\neq\emptyset\right)\) for some \(\vec{r}\in\left(0,1\right)^{N}\).

\smallskip

Put \(d\coloneq\mathbb{D}_{\vec{r}}\left(S\neq\emptyset\right)\). Since \(i\in S\) implies \(S\neq\emptyset\), admissibility gives
\[w_{i}=\mathbb{P}\left(i\in S\right)=\mathbb{D}_{\vec{r}}\left(i\in S\mid S\neq\emptyset\right)=\mfrac{r_{i}}{d},\quad i\in\left[N\right].\]
Thus \(\vec{r}=d\vec{w}\) and
\[d=\mathbb{D}_{d\vec{w}}\left(S\neq\emptyset\right)=1-\prod\nolimits_{i=1}^{N}\left(1-dw_{i}\right).\]
Since \(0<d<1\), the uniqueness in \cref{dfn:overfac} gives \(d=c_{\vec{w}}\), hence \(\mathbb{P}=\mathbb{D}_{c_{\vec{w}}\vec{w}}\left(\cdot\mid S\neq\emptyset\right)\).
\end{proof}

\section{An integral identity}\label{app:integral}

For \(\vec{r}=\left(r_{1},\dotsc,r_{N}\right)\in\left[0,1\right]^{N}\) and \(T\subseteq\left[N\right]\), recall the quantity \(p_{T}^{\vec{r}}\) defined in \eqref{eq:diskin-regions}. Let us establish that for every \(i\in\left[N\right]\) the following identities hold:
\begin{equation}\label{eq:integral}
\sum\nolimits_{\left\{T\subseteq\left[N\right]:i\in T\right\}}\mfrac{p_{T}^{\vec{r}}}{\left|T\right|}=r_{i}\int_{0}^{1}\prod\nolimits_{j\in\left[N\right]\setminus\{i\}}\left(1-r_{j}+r_{j}t\right)dt=r_{i}\sum\nolimits_{J\subseteq\left[N\right]\setminus\left\{i\right\}}\mfrac{\left(-1\right)^{\left|J\right|}}{\left|J\right|+1}\prod\nolimits_{j\in J}r_{j}.
\end{equation}

\smallskip

Fix \(i\in\left[N\right]\). In expanding the product \(\prod_{j\in\left[N\right]\setminus\{i\}}\left(1-r_{j}+r_{j}t\right)\), choose the term \(r_{j}t\) for the indices in a set \(J\) and the term \(1-r_{j}\) for the remaining indices. Thus
\[\prod\nolimits_{j\in\left[N\right]\setminus\{i\}}\left(1-r_{j}+r_{j}t\right)=\sum\nolimits_{J\subseteq\left[N\right]\setminus\left\{i\right\}}t^{\left|J\right|}\prod\nolimits_{j\in J}r_{j}\prod\nolimits_{j\in\left[N\right]\setminus\left(J\cup\{i\}\right)}\left(1-r_{j}\right).\]
Writing \(T=\left\{i\right\}\cup J\) and multiplying by \(r_{i}\), equations \eqref{eq:indep} and \eqref{eq:diskin-regions} give
\[r_{i}\prod\nolimits_{j\in\left[N\right]\setminus\{i\}}\left(1-r_{j}+r_{j}t\right)=\sum\nolimits_{\left\{T\subseteq\left[N\right]:i\in T\right\}}p_{T}^{\vec{r}}t^{\left|T\right|-1}.\]
Integrating from \(0\) to \(1\) and using \(\int_{0}^{1}t^{\left|T\right|-1}dt=1/\left|T\right|\) gives the first equality.

\smallskip

For the second equality, substitute \(u=1-t\) in the integral. Choosing \(-r_{j}u\) for the indices in \(J\) and \(1\) for the remaining indices gives
\[\prod\nolimits_{j\in\left[N\right]\setminus\{i\}}\left(1-r_{j}u\right)=\sum\nolimits_{J\subseteq\left[N\right]\setminus\left\{i\right\}}\left(-1\right)^{\left|J\right|}u^{\left|J\right|}\prod\nolimits_{j\in J}r_{j}.\]
Integrating term by term and using \(\int_{0}^{1}u^{\left|J\right|}du=1/\left(\left|J\right|+1\right)\) gives the second formula.

\section{Properties of the max-entropy rule}\label{app:properties}

\begin{prop}\label{prop:properties}
The max-entropy rule is efficient, association bounded, anonymous, strictly cross-association monotone, and strictly association monotone. The rule \(\left(\operatorname{ME}_{N}\right)_{N\geq1}\) is strictly population monotone and extends concede-and-divide: \(\operatorname{ME}_{2}=\operatorname{CD}\).
\end{prop}

The extension of concede-and-divide is verified in the proof of \cref{thm:shares}. For the standard properties we follow \citet{Thomson2019}: efficiency, called \emph{balance} there (\S3.1); association boundedness, called \emph{claims boundedness} there (\S3.4.1); anonymity (\S3.6). For the max-entropy rule, efficiency and anonymity were verified in the proof of \cref{thm:shares}, and association boundedness follows from \eqref{eq:alloceuiv}.

\smallskip

A rule \(R_{N}:\Omega_{N}\longrightarrow\left[0,1\right]^{N}\) is \emph{strictly cross-association monotone} if whenever \(\vec{w},\vec{w}^{\prime}\in\Omega_{N}\) satisfy \(w_{j}^{\prime}>w_{j}\) and \(w_{i}^{\prime}=w_{i}\) for every \(i\neq j\),
\[R_{N,i}\left(\vec{w}^{\prime}\right)<R_{N,i}\left(\vec{w}\right),\quad i\neq j\text{ with }w_{i}>0.\]
It is \emph{strictly association monotone} if, under the same conditions,
\[R_{N,j}\left(\vec{w}^{\prime}\right)>R_{N,j}\left(\vec{w}\right).\]
These are the strict versions of \emph{other-regarding claim monotonicity} and \emph{claims monotonicity} \citep[\S4.2]{Thomson2019}. A rule \(\left(R_{N}\right)_{N\geq1}\) is \emph{strictly population monotone} if for every \(\vec{w}\in\Omega_{N}\) and \(u\in\left(0,1\right]\),
\[R_{N+1,i}\left(\vec{w},u\right)<R_{N,i}\left(\vec{w}\right)\text{ for all }i\in\left[N\right]\text{ with }w_{i}>0.\]
This is the strict version of population monotonicity; see \citet[\S2.5]{Moulin2003} and \citet[\S12.1]{Thomson2019}.

\smallskip

The case \(\sum_{i=1}^{N}w_{i}=1\) is a routine check using \(\operatorname{ME}_{N}\left(\vec{w}\right)=\vec{w}\), the formula \eqref{eq:alloc-integral}, and efficiency. Assume then that \(\sum_{i=1}^{N}w_{i}>1\). For the max-entropy rule, these properties follow from a common observation. For every \(t\in\left(0,1\right)\), each factor \(1-c_{\vec{w}}w_{k}+c_{\vec{w}}w_{k}t\) in \eqref{eq:alloc-integral} is positive and strictly decreasing as \(c_{\vec{w}}w_{k}\) increases. By \cref{lem:overfac}(1), increasing \(w_{j}\) weakly increases all these parameters and strictly increases the parameter of \(j\). Thus, for every \(i\neq j\) with \(w_{i}>0\), the integrand in \eqref{eq:alloc-integral} strictly decreases while \(w_{i}\) remains positive. Null awards remain zero. This proves strict cross-association monotonicity.

\smallskip

Similarly, by \cref{lem:overfac}(2), adding an agent with association \(u>0\) weakly increases all the original parameters and introduces an additional factor
\[1-c_{\left(\vec{w},u\right)}u+c_{\left(\vec{w},u\right)}ut\in\left(0,1\right),\quad t\in\left(0,1\right).\]
Then the award of every original agent with positive association strictly decreases, proving strict population monotonicity. For strict association monotonicity, since \(w_{j}<1\), at least one other association is positive. The awards of all such agents strictly decrease, while null awards remain zero. Efficiency therefore gives a strict increase in the award of \(j\), since
\[\operatorname{ME}_{N,j}\left(\vec{w}\right)=1-\sum\nolimits_{i\neq j}\operatorname{ME}_{N,i}\left(\vec{w}\right).\]

\section{Supplementary calculations}
\label{app:comparison-formulas}

\subsection{Response to an increase of a claim}\label{app:suppwt}

We derive the formulas for \cref{exm:minimal-overlap}. For \(0\leq t\leq 1\) put \(\vec{w}\left(t\right)\coloneq\left(1,1/2,t\right)\). The proportional formula is verified directly to be
\[\operatorname{PROP}_{3,2}\left(\vec{w}\left(t\right)\right)=\mfrac{1}{3+2t}.\]

\smallskip

For the max-entropy rule, \(\operatorname{ME}_{3}\left(\vec{w}\left(t\right)\right)=\operatorname{Disk}_{3}\left(\vec{w}\left(t\right)\right)\) by \eqref{eq:me-extends-diskin} as \(\vec{w}\left(t\right)\in\Omega_{3}^{\mathrm{tot}}\). The second expression in \eqref{eq:diskin-integral} then gives
\[\operatorname{ME}_{3,2}\left(\vec{w}\left(t\right)\right)=\mfrac{1}{2}\Big(1-\mfrac{1+t}{2}+\mfrac{t}{3}\Big)=\mfrac{1}{4}-\mfrac{t}{12}.\]

\smallskip

For the Talmud rule, we use the formula in \citet[\S2.1.6, p.~34]{Thomson2019}. Since \(\sum_{i=1}^{3}w_{i}\left(t\right)/2\geq1\) precisely when \(t\geq1/2\), it gives
\[\operatorname{Talmud}_{3,2}\left(\vec{w}\left(t\right)\right)=\begin{cases}1/2-\min\left\{1/4,\lambda\right\}&0\leq t<1/2\\\min\left\{1/4,\lambda\right\}&1/2\leq t\leq1\end{cases},\]
where \(\lambda\geq0\) is such that \(\sum_{i=1}^{3}\min\left\{w_{i}\left(t\right)/2,\lambda\right\}\) equals \(1/2+t\) in the first case and \(1\) in the second. If \(\lambda<1/4\), this sum is less than \(1/2+t/2\leq\min\left\{1/2+t,1\right\}\), which is impossible in both cases. Hence \(\lambda\geq1/4\), and therefore
\[\operatorname{Talmud}_{3,2}\left(\vec{w}\left(t\right)\right)=\mfrac{1}{4},\quad 0\leq t\leq1.\]

\smallskip

For the random arrival rule, we use the formula in \citet[\S2.1.8, p.~38]{Thomson2019}. There are six arrival orders. The second claimant receives \(1/2\) when he arrives first (in two orders), nothing when he arrives after the first claimant (in three orders), and \(\min\left\{1/2,1-t\right\}\) otherwise (in one order). Hence
\[\operatorname{RA}_{3,2}\left(\vec{w}\left(t\right)\right)=\mfrac{1+\min\left\{1/2,1-t\right\}}{6}=\begin{cases}1/4&0\leq t\leq1/2\\\left(2-t\right)/6& 1/2<t\leq1\end{cases}.\]

\smallskip

For the minimal overlap rule, since the first claim is total, the claims are nested \citep[Prop.~2.1, Case~1, p.~41]{Thomson2019}: they are \(\left[0,1\right]\), \(\left[0,1/2\right]\), \(\left[0,t\right]\). The second claimant shares \(\left[0,\min\left\{t,1/2\right\}\right]\) with both other claimants and the rest of his claim with the first claimant only. Hence
\[\operatorname{MO}_{3,2}\left(\vec{w}\left(t\right)\right)=\mfrac{\min\left\{t,1/2\right\}}{3}+\mfrac{1/2-\min\left\{t,1/2\right\}}{2}=\begin{cases}1/4-t/6&0\leq t\leq1/2\\1/6&1/2<t\leq1\end{cases}.\]

\smallskip

For the average-of-awards rule, since the rule is anonymous, we can reorder the claims by their size. For \(0\leq t\leq1/2\), the two smaller claims sum to \(1/2+t\leq1\), and the largest claim equals the estate, so by \citet[eq.~(2)]{Miras2022},
\[\operatorname{AA}_{3}\left(\vec{w}\left(t\right)\right)=\Big(\mfrac{3}{4}-\mfrac{t}{2},\mfrac{1}{4},\mfrac{t}{2}\Big),\quad 0\leq t\leq1/2.\]
For \(1/2<t\leq1\), the second claim is the smallest, and the estate is at most half the total claim, as required in \citet[Lemma~5.2]{Miras2022}. Applying its formula for \(\operatorname{AA}_{1}\) with \(\left(d_{1},d_{2},d_{3}\right)=\left(1/2,t,1\right)\) and \(E=1\in\left[d_{2},\min\left\{d_{1}+d_{2},d_{3}\right\}\right]\) gives
\[\operatorname{AA}_{3,2}\left(\vec{w}\left(t\right)\right)=\mfrac{1/2-\left(1-t\right)^{3}}{3\big(3/4-\left(1-t\right)^{2}\big)}=\mfrac{4t^{3}-12t^{2}+12t-2}{3\left(8t-4t^{2}-1\right)},\quad 1/2<t\leq1.\]

\subsection{Full-information benchmark compatibility}\label{app:suppfinf}

Recall \cref{dfn:bench-comp} of full-information benchmark compatibility. For two claimants, the admissible coupling is unique and gives concede-and-divide, and hence full-information benchmark compatibility forces \(R_{2}=\operatorname{CD}\). Among the rules considered in \S\ref{sct:comparison}, the max-entropy rule is full-information benchmark compatible by definition, and so is the minimal overlap rule: it chooses an admissible arrangement of the claims and divides each region equally among its claimants \citep[\S2.1.9]{Thomson2019}.

\smallskip

We will show that all other rules considered in \S\ref{sct:comparison} are not full-information benchmark compatible. For the proportional rule, this is immediate since it does not extend concede-and-divide. For the Talmud and the average-of-awards rules, look at \(\vec{w}\coloneq\vec{w}\left(1\right)=\left(1,1/2,1\right)\). For every admissible coupling \(\mathbb{P}\in\mathcal{P}_{3}\left(\vec{w}\right)\) we have \(1,3\in S\) almost surely and \(\mathbb{P}\left(2\in S\right)=1/2\), hence
\[\Phi_{3,2}\left(\mathbb{P}\right)=\mathbb{E}_{\mathbb{P}}\Big[\mfrac{\mathbf{1}_{\left\{2\in S\right\}}}{\left|S\right|}\Big]=\mathbb{P}\left(2\in S\right)/3=1/6.\]
However, the formulas in \S\ref{app:suppwt} for \(t=1\) give
\[\operatorname{Talmud}_{3,2}\left(\vec{w}\right)=1/4\quad\text{and}\quad\operatorname{AA}_{3,2}\left(\vec{w}\right)=2/9.\]
Thus these two rules are not full-information benchmark compatible.

\smallskip

The following will be used to test full-information benchmark compatibility for the remaining rules.

\begin{lem}\label{lem:ficomp}
If a rule \(\left(R_{N}\right)_{N\geq 1}\) is full-information benchmark compatible, then for every \(\vec{w}\in\Omega_{N}\),
\begin{equation}\label{eq:bound}
R_{N,i}\left(\vec{w}\right)\geq w_{i}-\mfrac{1}{2}\Big(\sum\nolimits_{k=1}^{N}w_{k}-1\Big),\quad i\in\left[N\right],
\end{equation}
and
\begin{equation}\label{eq:Jens}
R_{N,i}\left(\vec{w}\right)\geq1/\sum\nolimits_{k=1}^{N}w_{k},\quad i\in\left[N\right],\,w_{i}=1.
\end{equation}
\end{lem}

\begin{proof}
Since \(\left|S\right|\geq1\) we have the general inequality
\[\mathbf{1}_{\left\{i\in S\right\}}\cdot\mfrac{\left|S\right|-1}{\left|S\right|}\leq\mfrac{\left|S\right|-1}{2},\quad i\in\left[N\right].\]
Then for every \(\mathbb{P}\in\mathcal{P}_{N}\left(\vec{w}\right)\), taking expectations, and using \eqref{eq:alloceuiv} and \(\mathbb{E}_{\mathbb{P}}\left[\left|S\right|\right]=\sum_{k=1}^{N}w_{k}\), we get
\[w_{i}-\Phi_{N,i}\left(\mathbb{P}\right)=\mathbb{E}_{\mathbb{P}}\Big[\mathbf{1}_{\left\{i\in S\right\}}\Big(1-\mfrac{1}{\left|S\right|}\Big)\Big]\leq\mfrac{1}{2}\mathbb{E}_{\mathbb{P}}\big[\left|S\right|-1\big]=\mfrac{1}{2}\Big(\sum\nolimits_{k=1}^{N}w_{k}-1\Big).\]
If \(w_{i}=1\), then \(i\in S\) almost surely by admissibility, and Jensen's inequality gives
\[\Phi_{N,i}\left(\mathbb{P}\right)=\mathbb{E}_{\mathbb{P}}\Big[\mfrac{1}{\left|S\right|}\Big]\geq\mfrac{1}{\mathbb{E}_{\mathbb{P}}\left[\left|S\right|\right]}=\mfrac{1}{\sum_{k=1}^{N}w_{k}}.\]
By full-information benchmark compatibility, \(R_{N,i}\left(\vec{w}\right)=\Phi_{N,i}\left(\mathbb{P}\right)\) for some \(\mathbb{P}\in\mathcal{P}_{N}\left(\vec{w}\right)\), so these two inequalities give \eqref{eq:bound} and \eqref{eq:Jens}.
\end{proof}

While the random stakes rule follows equal division in every round, this by itself does not ensure full-information benchmark compatibility. For three claimants and \(\left\{i,j,k\right\}=\left\{1,2,3\right\}\), the first-round award of claimant \(i\) is given in \citet[p.~44, proof of Prop.~2.2]{Thomson2019}, with \(E=1\) and \(p_{i}=w_{i}\):
\begin{equation}\label{eq:rs-first-round}
a_{i}\coloneq w_{i}\left(1-w_{j}\right)\left(1-w_{k}\right)
+\mfrac{w_{i}w_{j}\left(1-w_{k}\right)+w_{i}w_{k}\left(1-w_{j}\right)}{2}
+\mfrac{w_{i}w_{j}w_{k}}{3}=w_{i}\big(1-\mfrac{w_{j}+w_{k}}{2}+\mfrac{w_{j}w_{k}}{3}\big).
\end{equation}
The three terms in the expanded expression correspond respectively to units claimed only by \(i\), by \(i\) and exactly one other claimant, and by all three claimants. Thus \(a_{i}=\sum_{\left\{T\subseteq\left[3\right]:i\in T\right\}}p_{T}^{\vec{w}}/\left|T\right|\), and the second equality is \eqref{eq:integral} with \(\vec{r}=\vec{w}\). The remaining part has size \(\mathbb{D}_{\vec{w}}\left(S=\emptyset\right)=p_{\emptyset}^{\vec{w}}=\left(1-w_{1}\right)\left(1-w_{2}\right)\left(1-w_{3}\right)\).

\begin{exm}\label{exm:random-stakes}
For \(0<t<1/6\) set \(\vec{w}\left(t\right)\coloneq\left(t,t,1-t/2\right)\in\Omega_{3}\). By \eqref{eq:bound}, every full-information benchmark compatible rule \(\left(R_{N}\right)_{N\geq1}\) must satisfy
\[R_{3,3}\left(\vec{w}\left(t\right)\right)\geq1-\mfrac{t}{2}-\mfrac{1}{2}\cdot\mfrac{3t}{2}=1-\mfrac{5t}{4}.\]
Let us calculate the random stakes rule for \(\vec{w}\left(t\right)\). By \eqref{eq:rs-first-round}, in the first round the third claimant receives
\[a_{3}=\big(1-\mfrac{t}{2}\big)\big(1-t+\mfrac{t^{2}}{3}\big),\]
and the remaining part has size
\[r\coloneq p_{\emptyset}^{\vec{w}\left(t\right)}=t\left(1-t\right)^{2}/2.\]
The remaining claim of each of the first two claimants is
\[t-a_{1}=\mfrac{t}{2}-\mfrac{t^{2}}{12}+\mfrac{t^{3}}{6}=r+\mfrac{t^{2}\left(11-4t\right)}{12}>r,\]
and that of the third claimant is
\[1-\mfrac{t}{2}-a_{3}=\big(1-\mfrac{t}{2}\big)\big(t-\mfrac{t^{2}}{3}\big)>r.\]
The claims are truncated at the remaining part \citep[\S2.1.10, p.~43]{Thomson2019}, so in the second round every claimant claims the entire remainder, so it is divided equally. Hence we get
\[\operatorname{RS}_{3,3}\left(\vec{w}\left(t\right)\right)=a_{3}+\mfrac{r}{3}=1-\mfrac{4t}{3}+\mfrac{t^{2}}{2}=1-\mfrac{5t}{4}-\mfrac{t\left(1-6t\right)}{12}<1-\mfrac{5t}{4}.\]
Therefore, the random stakes rule is not full-information benchmark compatible.
\end{exm}

We finally show that the random arrival rule is not full-information benchmark compatible.

\begin{exm}\label{exm:random-arriv}
Let \(\vec{w}=\left(1,1/2,1/2,1/2\right)\). By \eqref{eq:Jens}, since \(w_{1}=1\) and \(\sum_{k=1}^{4}w_{k}=5/2\), every full-information benchmark compatible rule \(\left(R_{N}\right)_{N\geq 1}\) must satisfy
\[R_{4,1}\left(\vec{w}\right)\geq\mfrac{2}{5}.\]
By the formula in \citet[\S2.1.8, p.~38]{Thomson2019}, the first claimant receives \(1\), \(1/2\), \(0\), \(0\) when he arrives first, second, third, fourth, respectively. Each position has probability \(1/4\), so we get
\[\operatorname{RA}_{4,1}\left(\vec{w}\right)=\mfrac{1}{4}\Big(1+\mfrac{1}{2}\Big)=\mfrac{3}{8}<\mfrac{2}{5}.\]
Hence the random arrival rule is not full-information benchmark compatible either.
\end{exm}

\addtocontents{toc}{\protect\setcounter{tocdepth}{1}}

\begin{table}[ht]
\captionsetup{font=small}
\centering
\small
\begin{tabular}{lp{0.70\textwidth}}
\toprule
Notation & Meaning \\
\midrule
\(\Omega_{N}\) & Vectors \(\vec{w}\in\left[0,1\right]^{N}\) with \(\sum_{i=1}^N w_{i}\geq1\).\\
\(\Omega_{N}^{\mathrm{tot}}\) & The vectors \(\vec{w}\in\Omega_{N}\) with \(\max_{1\leq i\leq N}w_{i}=1\).\\
\(\left(R_{N}\right)_{N\geq1}\) & A general rule, with \(R_{N}:\Omega_{N}\longrightarrow\left[0,1\right]^{N}\).\\
\(\left(a,\vec{w}_{-j}\right)\) & The vector \(\vec{w}\) with \(a\) in its \(j^{\mathrm{th}}\) coordinate; \(\vec{w}_{-j}\) is the vector with \(w_{j}\) removed.\\
\(\mathcal{S}_{N}\) & The family of nonempty subsets of \(\left[N\right]=\{1,\dotsc,N\}\).\\
\(S\) & The set of agents associated with a random unit.\\
\(X_{i}\) & The association indicator \(X_{i}=\mathbf{1}_{\{i\in S\}}\), with \(\mathbb{P}\left(X_{i}=1\right)=w_{i}\) for all \(\mathbb{P}\in\mathcal{P}_{N}\left(\vec{w}\right)\).\\
\(\mathcal{P}_{N}\left(\vec{w}\right)\) & The set of admissible couplings of an association vector \(\vec{w}\).\\
\(\mathbb{D}_{\vec{w}}\) & The independent coupling of an association vector \(\vec{w}\).\\
\(p_{T}=\mathbb{P}\left(T\right)\) & The size of the region associated with the members of \(T\) under \(\mathbb{P}\).\\
\(p_{T}^{\vec{w}}\) & The weights of the independent coupling, \(p_{T}^{\vec{w}}=\mathbb{D}_{\vec{w}}\left(T\right)\).\\
\(\mathbb{P}_{\vec{w}}^{\ast}\) & The max-entropy element of \(\mathcal{P}_{N}\left(\vec{w}\right)\).\\
\(c_{\vec{w}}\) & The overlap factor of \(\vec{w}\) (\cref{dfn:overfac}).\\
\(\Phi_{N}\left(\mathbb{P}\right)\) & The vector of shares obtained by equal division under the overlap model \(\mathbb{P}\).\\
\(v_{\mathbb{P}}\) & The coalitional game \(v_{\mathbb{P}}\left(A\right)=\mathbb{P}\left(S\subseteq A\right)\).\\
\(\operatorname{Disk}_{N}\) & Diskin's rule (\cref{dfn:diskin-rule}).\\
\(\operatorname{ME}_{N}\) & The max-entropy rule, \(\operatorname{ME}_{N}\left(\vec{w}\right)=\Phi_{N}\left(\mathbb{P}_{\vec{w}}^{\ast}\right)\).\\
\(\operatorname{CD}\) & Concede-and-divide for two agents.\\
\(\operatorname{RC}_{N}\) & O'Neill's random claims rule \eqref{eq:rc-rule}.\\
\(\operatorname{RS}_{N}\) & The random stakes rule (\S\ref{sct:randomclaims}).\\
\bottomrule
\end{tabular}
\caption{Main notation.}
\label{tab:notation}
\end{table}

\section*{Use of AI}

The problem formulation, the resource consistency axiom, and the identification of the rule with the maximum-entropy coupling are solely due to the authors. Some of the proofs were developed by the authors with assistance from AI tools (Claude Opus 5.5 and ChatGPT Sol 6.2). These tools were also used to search the literature, check for mathematical and language errors, write the TikZ code for \cref{fig:secawards}, and find and work out examples~\ref{exm:minimal-overlap}, \ref{app:suppwt}, \ref{exm:random-stakes}, \ref{exm:random-arriv}. The manuscript was written only by the authors, who verified the proofs and computations and are responsible for its content.

\section*{Acknowledgments}

The authors thank Yosef Abraham and Michael Levin for a stimulating discussion. N. Avraham-Re'em was supported by the ISF (grant No. 3423/24) and an Aly Kaufman Fellowship at the Technion.

\bibliographystyle{chicago}
\bibliography{References}

\end{document}